\documentclass[prd,aps,twocolumn,nofootinbib,preprintnumbers]{revtex4}
\usepackage{graphicx,amsmath}
\usepackage{longtable,dcolumn}
\newcolumntype{d}[1]{D{.}{.}{#1}}

\begin{document}


\preprint{KEK-TH-1900}
\title{
Gravitational waves from bubble collisions: analytic derivation
}
\renewcommand{\thefootnote}{\alph{footnote}}

\author{
Ryusuke Jinno
and
Masahiro Takimoto}

\affiliation{
Theory Center, High Energy Accelerator Research Organization (KEK), 1-1 Oho, Tsukuba, Ibaraki 305-0801, Japan\\
}

\begin{abstract}

We consider gravitational wave production by bubble collisions
during a cosmological first-order phase transition.
In the literature, such spectra have been estimated by simulating the bubble dynamics, 
under so-called thin-wall and envelope approximations in a flat background metric.
However, we show that, within these assumptions, 
the gravitational wave spectrum can be estimated in an analytic way.
Our estimation is based on the observation that
the two-point correlator of the energy-momentum tensor $\langle T(x)T(y)\rangle$
can be expressed analytically
under these assumptions. 
Though the final expressions for the spectrum contain 
a few integrations that cannot be calculated explicitly, 
we can easily estimate it numerically.
As a result, it is found that the most of the contributions to the spectrum come from
single-bubble contribution to the correlator,
and in addition the fall-off of the spectrum at high frequencies is found to be proportional to $f^{-1}$.
We also provide fitting formulae for the spectrum.

\end{abstract}

\maketitle

\section{Introduction}

Gravitational waves (GWs) are one of the promising tools to probe the early universe.
They provide a unique way to search for 
inflationary quantum fluctuations~\cite{Starobinsky:1979ty}, 
preheating~\cite{Khlebnikov:1997di},
topological defects~\cite{Vilenkin:2000jqa,Gleiser:1998na}, 
and cosmic phase transitions (PTs)~\cite{Witten:1984rs,Hogan:1986qda}.
Especially, first-order PTs in the early universe have been attracted us
because of their relation to high-energy physics beyond the standard model (SM), 
and in fact various extensions of the SM have been shown to
predict first-order PTs with a large amount of GWs~\cite{
Espinosa:2008kw,Ashoorioon:2009nf,Das:2009ue,Sagunski:2012pzo,Kakizaki:2015wua,Jinno:2015doa,Apreda:2001tj,Apreda:2001us,Jaeckel:2016jlh,Huber:2015znp,Leitao:2015fmj,Huang:2016odd,Dev:2016feu,Hashino:2016rvx,Jinno:2016knw}.
On the observational side,
ground-based GW experiments like KAGRA~\cite{Somiya:2011np},
VIRGO~\cite{TheVirgo:2014hva} and Advanced LIGO~\cite{Harry:2010zz}
are now in operation, 
and space interferometers such as
eLISA~\cite{Seoane:2013qna}, BBO~\cite{Harry:2006fi} 
and DECIGO~\cite{Seto:2001qf}
have been proposed.
Given that there is a growing possibility of their detecting 
GWs from cosmological sources in the near future,
it would be worth reconsidering the theoretical predictions of GWs from first-order PTs.

First-order PTs proceed via the nucleation of bubbles, their expansion,
collision and thermalization into light particles,
and GWs are produced during this process.
In the transition process,
some of the released energy goes into heating up the plasma,
while the rest is carried by the scalar field configuration (bubble wall) 
and/or the bulk motion of the surrounding fluid.
Gravitational wave production by such localized structure of energy around the walls 
has been calculated by numerical simulations in the literature
with so-called thin-wall and envelope approximations~\cite{Kosowsky:1991ua,Kosowsky:1992rz,Kosowsky:1992vn,Kamionkowski:1993fg}\footnote{
It is important to go beyond these approximations, 
especially when the bulk motion of the fluid dominates the released energy.
In fact, it has been pointed out that the bulk motion can be a long-lasting GW source 
as sound waves~\cite{Hindmarsh:2013xza,Giblin:2014qia,Hindmarsh:2015qta}.
}.
It has been shown that these approximations are valid especially 
when the energy of bubbles is dominated by the scalar field configuration~\cite{Kosowsky:1991ua,Weir:2016tov},
and the latest result along this approach is found in Ref.~\cite{Huber:2008hg}.
Analytic approaches have also been taken
with some ansatz for correlator functions~\cite{Caprini:2007xq,Caprini:2009fx}.

In this paper, 
we take an approach
based on the evaluation of the correlation function 
of the energy-momentum tensor
$\langle T(x) T(y) \rangle$~\cite{Caprini:2007xq},
which is the only ingredient to obtain the spectrum.
We point out that,
under thin-wall and envelope approximations and in a flat background,
this two-point correlator has a rather simple analytic expression and, 
as a result, the GW spectrum can also be expressed analytically.
Though the final expression for the spectrum contains two remaining integrations,
they can easily be estimated numerically.
Our approach is not only free from statistical errors inherent to numerical simulations,
but also enables us to specify the most effective bubble-wall configuration 
to the GW spectrum.
At the current stage, 
our results are most relevant to strong phase transitions like near-vacuum ones,
since the neglected effects such as 
the finite width of the bubble walls
and/or the localized structure of the energy-momentum tensor remaining after collisions
can be important when the scalar field is strongly coupled 
to the thermal plasma~\cite{Hindmarsh:2013xza,Giblin:2014qia,Hindmarsh:2015qta}~\footnote{
In addition, turbulent effects can contribute sizably to the 
GW spectrum~\cite{Kamionkowski:1993fg,Caprini:2006jb,Gogoberidze:2007an,Caprini:2009yp}.
}.
However, our method is extendable 
to the calculations without the envelope approximation~\cite{JT},
and such studies would be important in understanding 
how the localized structure after bubble collisions sources GWs.

The organization of the paper is as follows.
In Sec.~\ref{sec:bi} 
we first make clear our assumptions in estimating the GW spectrum,
i.e. thin-wall and envelope approximations,
and then introduce basic ingredients such as 
the evolution equation and power spectrum of GWs.
In Sec.~\ref{sec:ana} we present 
analytic expressions for the GW spectrum.
Since two integrations cannot be performed explicitly,
we evaluate them numerically in Sec.~\ref{sec:num}.
We generalize our result to finite velocity case 
in Sec.~\ref{sec:vel},
and finally summarize in Sec.~\ref{sec:con}.

\section{Basic ingredients}
\label{sec:bi}

In this section we summarize basic ingredients for the calculation of GW spectrum.
We first make clear the assumption and approximations used in the paper.
We also explain the GW power spectrum around the time of sourcing from bubble collisions,
and then show how to obtain the present spectrum.

\subsection{Assumptions and approximations}

\subsubsection{Thin wall and envelope approximation}

In this subsection, 
we introduce the key assumptions to characterize the energy momentum tensor around the bubble wall,
namely thin-wall and envelope approximations.

\begin{figure}
\begin{center}
\includegraphics[width=0.7\columnwidth]{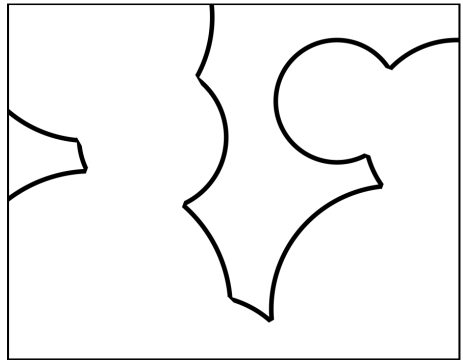}
\caption {\small
Rough sketch of how the phase transition looks with the envelope approximation.
Collided walls are neglected as a source of GWs,
and all spacial points are passed by bubble walls only once.
}
\label{fig:Envelope}
\end{center}
\end{figure}

First, we introduce the thin-wall approximation,
where all the energy of the bubble is assumed to be concentrated on the bubble wall with an infinitesimal width.
We introduce the infinitesimal wall width $l_B$ for computational simplicity.
The energy momentum tensor $T^B$ of the uncollided wall of a single bubble
nucleated at $x_N \equiv (t_N, \vec{x}_N)$ can be written as 
\begin{align}
T_{ij}^B(x)
&= \rho(x)\widehat{(x - x_N)}_i\widehat{(x - x_N)}_j,
\label{eq:TB}
\end{align}
with
\begin{align}
\rho(x)
&=
\left\{
\begin{array}{cc}
\displaystyle 
\frac{4\pi}{3} r_B(t)^3 
\frac{\kappa\rho_0}{4\pi r_B(t)^2 l_B}
&
r_B(t) < |\vec{x} - \vec{x}_N| < r'_B(t) \\
0
& 
{\rm otherwise}
\end{array}
\right. 
\label{eq:rho}
\end{align}
and 
\begin{align}
r_B(t)
&= v(t - t_N), 
\;\;\;
r'_B(t)
= r_B(t) + l_B.
\end{align}
Here $x \equiv (t,\vec{x})$, 
the hat on the vector $\hat{\bullet}$ indicates the unit vector in the direction of $\vec{\bullet}$,
$v$ is the bubble wall velocity,
and $\rho_0$ represents the energy density released by the transition\footnote{
Though the corresponding quantity is latent heat and not energy density
in thermal environment,
we use the word ``energy density" throughout the paper, 
since 
}.
Also, $\kappa$ indicates the efficiency factor,
which determines the fraction of the released energy density 
which is transformed into 
the energy density localized around the wall\footnote{
This corresponds to the energy density of the bulk fluid around the wall
when the bubble wall reaches a terminal velocity, 
while it is regarded as the energy density of the wall itself
when the scalar field carries most of the energy.
In the former case with so-called Jouguet detonation,
the efficiency factor is related to the parameter $\alpha$ introduced later~\cite{Steinhardt:1981ct}.
}~\cite{Kamionkowski:1993fg}.
In addition, the Latin indices run over $1, 2, 3$ throughout the paper.
Second, we assume that 
the energy momentum tensor of the bubble walls vanishes once they collide with others.
In the literature this is called envelope approximation,
whose validity in bubble collisions is confirmed in \textit{e.g.} Ref.~\cite{Kosowsky:1991ua}.
See Fig.~\ref{fig:Envelope} for a rough sketch of this approximation.
These two assumptions make the calculation of the GW spectrum rather simple, as we will see later.
Also, we regard the model-dependent quantities 
$v$, $\rho_0$ and $\kappa$ as free parameters constant in time.

\subsubsection{Transition rate}

We assume that the bubble nucleation rate 
per unit time and volume can be written in the following form:
\begin{align}
\Gamma(t)
&= \Gamma_*e^{\beta(t - t_*)},
\label{eq:Gamma_beta}
\end{align}
where $t_*$ indicates some fixed time typically around the transition time,
$\Gamma_*$ is the nucleation rate at $t = t_*$, and $\beta$ is assumed to be a constant.
This parameter $\beta$ is often calculated with the instanton method
from underlying models~\cite{Linde:1977mm,Linde:1981zj}, 
and the typical time span of the phase transition is given by $\delta t\sim \beta^{-1}$.
We also assume that the phase transition completes in a short period 
compared to the Hubble time, \textit{i.e.} $\beta/H\gg1$, which typically holds 
for thermal phase transitions~\cite{Kamionkowski:1993fg}.

\subsection{GW power spectrum around the transition time}

In the following we express the GW spectrum 
in terms of the correlator of the energy-momentum tensor,
following Ref.~\cite{Caprini:2007xq}.

\subsubsection{Equation of motion and its solution}

In this paper we consider GWs sourced by the first order phase transition 
completed in a short period compared to the Hubble time.
In such cases the background metric is well approximated by the Minkowski one.
Including tensor perturbations, we write the metric as
\begin{align}
ds^2
&= - dt^2 + (\delta_{ij}+2h_{ij})dx^idx^j.
\end{align}
The tensor perturbations satisfy the transverse and traceless condition
$h_{ii} = \partial_j h_{ij} = 0$ and obey the following evolution equation
\begin{align}
\ddot{h}_{ij}(t,\vec{k})+k^2h_{ij} (t,\vec{k}) 
&= {8\pi G } \Pi_{ij}(t,\vec{k}),
\label{eq:hEOM}
\end{align}
where $G$ is the Newton constant and 
$\bullet (t,\vec{k})$ indicates a Fourier mode of the corresponding object
with $\vec{k}$ being the wave vector.
We take the convention for Fourier transformation to be 
$\int d^3x \; e^{i\vec{k}\cdot\vec{x}}$ and $\int d^3k/(2\pi)^3 \; e^{-i\vec{k}\cdot\vec{x}}$.
The source term $\Pi_{ij}$ during the phase transition is given by
the transverse and traceless projection of the energy momentum tensor 
\begin{align}
\Pi_{ij}(t,\vec{k})
&= K_{ij,kl}(\hat{k}) T_{kl}(t,\vec{k}),
\label{eq:PiKT}
\end{align}
with $T_{ij}$ being the energy momentum tensor,
and $K_{ij,kl}$ being the projection
\begin{align}
K_{ij,kl}(\hat{k})
&= P_{ik}(\hat{k})P_{jl}(\hat{k}) - \frac{1}{2}P_{ij}(\hat{k})P_{kl}(\hat{k}), 
\label{eq:kk}
\\
P_{ij}(\hat{k})
&\equiv \delta_{ij}-\hat{k}_i\hat{k}_j.
\end{align}
We assume that the source term is effective from $t_{\rm start}$ to $t_{\rm end}$,
and we set $t_{\rm start/end}\rightarrow \mp \infty$
at the end of calculation\footnote{
Since the transition completes in a short period $\delta t\sim \beta^{-1} \ll H^{-1}$,
and GWs are emitted only during this period,
this procedure is expected not to affect the result.
}.

The solution of Eq.~(\ref{eq:hEOM}) is formally written in terms of the Green function $G_k$ 
satisfying $G_k(t,t) = 0$ and $\partial G_k(t,t')/\partial t |_{t = t'} = 1$ as
\begin{align}
h_{ij}(t,\vec{k})
&= 8\pi G
\int_{t_{\rm start}}^t dt' \;
G_k(t,t') \Pi_{ij}(t',\vec{k})
\;\;\;\;\;\;
t < t_{\rm end},
\label{eq:hsol_G}
\end{align}
where $G_k(t,t') = \sin (k(t - t'))/k$.
For $t > t_{\rm end}$, matching condition at $t = t_{\rm end}$ gives
\begin{align}
h_{ij}(t,\vec{k})
&= A_{ij}(\vec{k}) \sin(k(t - t_{\rm end})) + B_{ij}(\vec{k}) \cos(k(t - t_{\rm end})),
\label{eq:hsol}
\end{align}
with coefficients
\begin{align}
A_{ij}(\vec{k})
&= \frac{8\pi G}{k}\int_{t_{\rm start}}^{t_{\rm end}}dt \;
\cos(k(t_{\rm end} - t)) \Pi_{ij}(t,\vec{k}), \\
B_{ij}(\vec{k})
&= \frac{8\pi G}{k}\int_{t_{\rm start}}^{t_{\rm end}}dt \;
\sin(k(t_{\rm end} - t')) \Pi_{ij}(t,\vec{k}).
\end{align}

\subsubsection{Power spectrum}

Next we express the GW spectrum using Eq.~(\ref{eq:hsol}).
We define the equal-time correlator of the GWs by
\begin{align}
\langle \dot{h}_{ij}(t,\vec{k}) \dot{h}_{ij}^*(t,\vec{q}) \rangle
&= (2\pi)^3 \delta^{(3)}(\vec{k} - \vec{q}) P_{\dot{h}}(t,k),
\end{align}
and also define the unequal-time correlator of the source term by
\begin{align}
\langle \Pi_{ij}(t_x,\vec{k})\Pi^*_{ij}(t_y,\vec{q}) \rangle
&= (2\pi)^3 \delta^{(3)}(\vec{k} - \vec{q})\Pi(t_x,t_y,k).
\label{eq:PiPi}
\end{align}
Here the angular bracket denotes taking an ensemble average.
Note that the $(2\pi)^3 \delta^{(3)}(\vec{k} - \vec{q})$ in Eq.~(\ref{eq:PiPi}) 
appears due to the spacial homogeneity of the system.
In terms the original energy-momentum tensor,
the correlator $\Pi(t_x,t_y,k)$ is written as
\begin{align}
&\Pi(t_x,t_y,k) \nonumber \\
&= K_{ij,kl}(\hat{k})K_{ij,mn}(\hat{k})
\int d^3r \; e^{i \vec{k} \cdot \vec{r}} \langle T_{kl} T_{mn} \rangle (t_x,t_y,\vec{r}),
\label{eq:Pi}
\end{align}
where 
\begin{align}
\langle T_{kl} T_{mn} \rangle (t_x,t_y,\vec{r})
&\equiv \langle T_{kl}(t_x,\vec{x}) T_{mn}(t_y,\vec{y}) \rangle,
\end{align}
with $\vec{r} \equiv \vec{x} - \vec{y}$.
The L.H.S. depends only on $\vec{r}$ because of the spacial homogeneity.
Now let us consider the time $t > t_{\rm end}$.
Since the GWs and the source term are related with each other through Eq.~(\ref{eq:hsol}),
the power spectrum of $P_{\dot{h}}$ is written in terms of the source as
\begin{align}
&P_{\dot{h}}(t,k) \nonumber \\
&= 32\pi^2G^2
\int_{t_{\rm start}}^{t_{\rm end}} dt_x
\int_{t_{\rm start}}^{t_{\rm end}} dt_y \;
\cos(k(t_x - t_y))\Pi (t_x,t_y,k).
\label{eq:Pdoth}
\end{align}
Though we put the argument $t$ in the L.H.S., 
the R.H.S. does not depend on it 
because there is no source term for $t > t_{\rm end}$ 
and because we neglect the cosmic expansion.
Since the total energy density of GWs is given by
\begin{align}
\rho_{\rm GW}(t)
&= \frac{\langle \dot{h}_{ij}(t,\vec{x})\dot{h}_{ij}(t,\vec{x}) \rangle_T}{8\pi G},
\end{align}
with $\langle \cdots \rangle_T$ being the oscillation and ensemble average,
GW energy density per logarithmic frequency becomes
\begin{align}
&\Omega_{\rm GW}(t,k) \nonumber \\
&\equiv \frac{1}{\rho_{\rm tot}} \frac{d\rho_{\rm GW}}{d\ln k} \nonumber \\
&= \frac{2Gk^3}{\pi \rho_{\rm tot}}
\int_{t_{\rm start}}^{t_{\rm end}} dt_x
\int_{t_{\rm start}}^{t_{\rm end}} dt_y \;
\cos(k(t_x - t_y))\Pi (t_x,t_y,k).
\end{align}
with $\rho_{\rm tot}$ being the total energy density of the universe.
Now all we have to do is to estimate $\Pi(t_x,t_y,k)$, 
or the two-point function of the energy momentum tensor $\langle T(x)T(y) \rangle$.
Once the setup is defined, we can estimate this quantity analytically in principle.
In fact, as shown later,
this correlator $\langle T(x)T(y) \rangle$ can be expressed in an an analytical way
under the thin-wall and envelope approximations
(see Eqs.~(\ref{eq:KKTT_single}) and (\ref{eq:KKTT_double})).

For later convenience, we rewrite the expression for the GW spectrum as follows.
We define the parameter $\alpha$ as
\begin{align}
\alpha
&\equiv \frac{\rho_0}{\rho_{\rm rad}},
\;\;\;\;\;\;
\rho_{\rm tot}
= \rho_0 + \rho_{\rm rad},
\label{eq:alpha}
\end{align}
which characterizes the fraction of the released energy density to that of radiation.
Here $\rho_{\rm tot}$ and $\rho_{\rm rad}$ are the total and radiation energy density, respectively.
Using $\alpha$ thus defined, we have
\begin{align}
\Omega_{\rm GW}(t,k)
&= \kappa^2 \left(\frac{H_*}{\beta}\right)^2\left(\frac{\alpha}{1+\alpha}\right)^2
\Delta(k/\beta,v),
\label{eq:Omega_Delta}
\end{align}
where $\Delta$ is given by
\begin{align}
&\Delta(k/\beta, v) \nonumber \\
&= \frac{3}{8\pi G}\frac{\beta^2 \rho_{\rm tot}}{\kappa^2 \rho_0^2}\Omega_{\rm GW}(t,k) \nonumber \\
&= \frac{3}{4\pi^2}\frac{\beta^2k^3}{\kappa^2\rho_0^2}
\int_{t_{\rm start}}^{t_{\rm end}} dt_x
\int_{t_{\rm start}}^{t_{\rm end}} dt_y \;
\cos(k(t_x - t_y))\Pi (t_x,t_y,k).
\label{eq:Delta}
\end{align}
In deriving Eq.~(\ref{eq:Omega_Delta}) we have used the Friedmann equation 
$H_*^2 = (8\pi G/3)\rho_{\rm tot}$ with $H_*$ being the Hubble parameter at the transition time.
Note that the function $\Delta$ depends only on the combination $k/\beta$ and the wall velocity $v$, 
because the definition (\ref{eq:Omega_Delta}) factors out
$\kappa$, $\rho_0$ and $\rho_{\rm tot}$ dependence, and 
because $\Delta$ is a dimensionless quantity.

\subsection{GW power spectrum at present}

After produced, GWs are redshifted
during propagation towards the present time.
The relation between the scale factor 
just after the phase transition $a_*$ 
and at present $a_0$ is given by
\begin{align}
\frac{a_0}{a_*}
&= 8.0\times10^{-16}
\left( \frac{g_*}{100} \right)^{-1} \left(\frac{T_*}{100~\text{GeV}}\right)^{-1},
\end{align}
where $T_*$ denotes the temperature just after the phase transition, 
and $g_*$ indicates the total number of the relativistic degrees of freedom in the thermal bath
at temperature $T_*$.
The present frequency is obtained by redshifting as
\begin{align}
f
&=
f_*\left(\frac{a_*}{a_0}\right) \nonumber \\
&= 1.65 \times 10^{-5} {\rm Hz}
\left( \frac{f_*}{\beta} \right) \left( \frac{\beta}{H_*} \right)
\left( \frac{T_*}{10^2{\rm GeV}} \right)
\left( \frac{g_*}{100} \right)^{\frac{1}{6}}, 
\label{eq:f_present}
\end{align}
and the present GW amplitude is obtained from the fact that 
GWs are non-interacting radiation as
\begin{align}
&\Omega_{\rm GW}h^2 \nonumber \\
&=1.67\times 10^{-5} \left( \frac{g_*}{100}\right)^{-\frac{1}{3}}
\Omega_{\rm GW}h^2\bigl|_{t = t_{\rm end}}\nonumber \\
&=1.67\times 10^{-5}\kappa^2 \Delta \left( \frac{\beta}{H_*} \right)^{-2}
\left( \frac{\alpha}{1 + \alpha} \right)^2
\left( \frac{g_*}{100} \right)^{-\frac{1}{3}}.
\label{eq:Omega_present}
\end{align}

\section{Analytic expression}
\label{sec:ana}

The following sections are mainly devoted to the calculation of $\Delta$ 
(see Eq.~(\ref{eq:Delta})).
We first focus on
the case where the wall velocity is luminal, \textit{i.e.}, $v = c$,
since the final explanations become relatively simple in this case.
Generalization to $v \neq c$ is straightforward and done in Sec.~\ref{sec:vel}.

\begin{figure}
\begin{center}
\includegraphics[width=0.6\columnwidth]{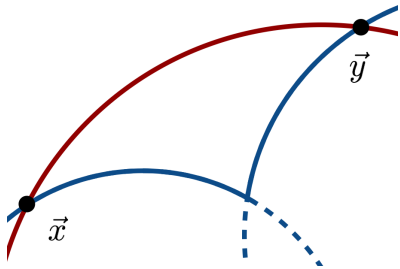}
\caption {\small
Schematic picture of single- and double-bubble contributions to the correlator $\langle T(x)T(y) \rangle$.
The red line (the one w/o dashed lines) corresponds to the wall of one single bubble,
while the blue line (the one w/ dashed lines) corresponds to intersecting two bubble walls.
The dashed lines are neglected in the envelope approximation.
Note that, though this figure shows $t_x = t_y$ case for simplicity, 
contributions from $t_x \neq t_y$ exist in the calculation of $\langle T(x) T(y) \rangle$.
}
\label{fig:SingleDouble}
\end{center}
\end{figure}

In the expression of the GW spectrum (\ref{eq:Delta}),
the only nontrivial quantity is the two-point correlator 
$\Pi(t_x,t_y,\vec{r})$ given by Eq.~(\ref{eq:Pi}).
If we can calculate this quantity, or equivalently 
the average of the product of the energy-momentum tensor
$\langle T_{ij} T_{kl} \rangle (t_x,t_y,\vec{r})$
with given $t_x$, $t_y$ and $\vec{r} = \vec{x} - \vec{y}$,
then we obtain the GW spectrum.
In the following we show that this is indeed possible.
For the energy momentum tensor to be nonzero
at $x = (t_x,\vec{x})$ and $y = (t_y,\vec{y})$ with $\vec{x} - \vec{y} = \vec{r}$, 
the following two conditions are necessary and sufficient:
\begin{itemize}
\item
No bubbles are nucleated inside the past light cones of $x$ and $y$.
\item
Bubble(s) are nucleated on the past light cones of $x$ and $y$,
so that bubble walls are passing through the spacial points $\vec{x}$ at time $t_x$ 
and $\vec{y}$ at time $t_y$. 
\end{itemize}
In order to understand the former condition, 
one needs to notice that any spacial point is passed by bubble walls 
only once in the envelope approximation (see Fig.~\ref{fig:Envelope}).
Then, if bubble(s) nucleate inside the past light cone of $x$ or $y$,
either of the spacial points $\vec{x}$ or $\vec{y}$ is already passed by bubble walls
before the evaluation time $t_x$ or $t_y$.
This makes it impossible for the energy-momentum tensor to be nonvanishing 
both at $x$ and $y$, and therefore we need the former condition.
On the other hand, the latter condition is necessary for bubble walls to be just passing through 
$\vec{x}$ and $\vec{y}$ at the evaluation time $t_x$ and $t_y$.
There are two possibilities for this condition: 
the bubble walls passing through $x$ and $y$ belong to one single nucleation point,
or to two different nucleation points.
We refer to these two as ``single-bubble" and ``double-bubble" contributions, respectively.
Fig.~\ref{fig:SingleDouble} shows a schematic picture of these two
contributions to the correlator $\langle T(x) T(y) \rangle$.
Here one may wonder why we consider the single-bubble contribution,
since it is well known that a spherical object do not radiate GWs.
The answer is that the single-bubble contribution takes into account 
the breaking of the original spherical symmetry of a bubble by collisions with others:
see Appendix~\ref{app:Saru} on this point.

In the following discussion, we first make our notation clear.
Then, after discussing the condition for
no bubble nucleation inside the past light cones,
we consider single- and double-bubble contributions separately.
The final expressions are Eqs.~(\ref{eq:Delta_single}) and (\ref{eq:Delta_double}),
and those who need only the final GW spectrum may skip to Sec.~\ref{sec:num}.

\subsection{Notations}

\begin{figure}
\begin{center}
\includegraphics[width=\columnwidth]{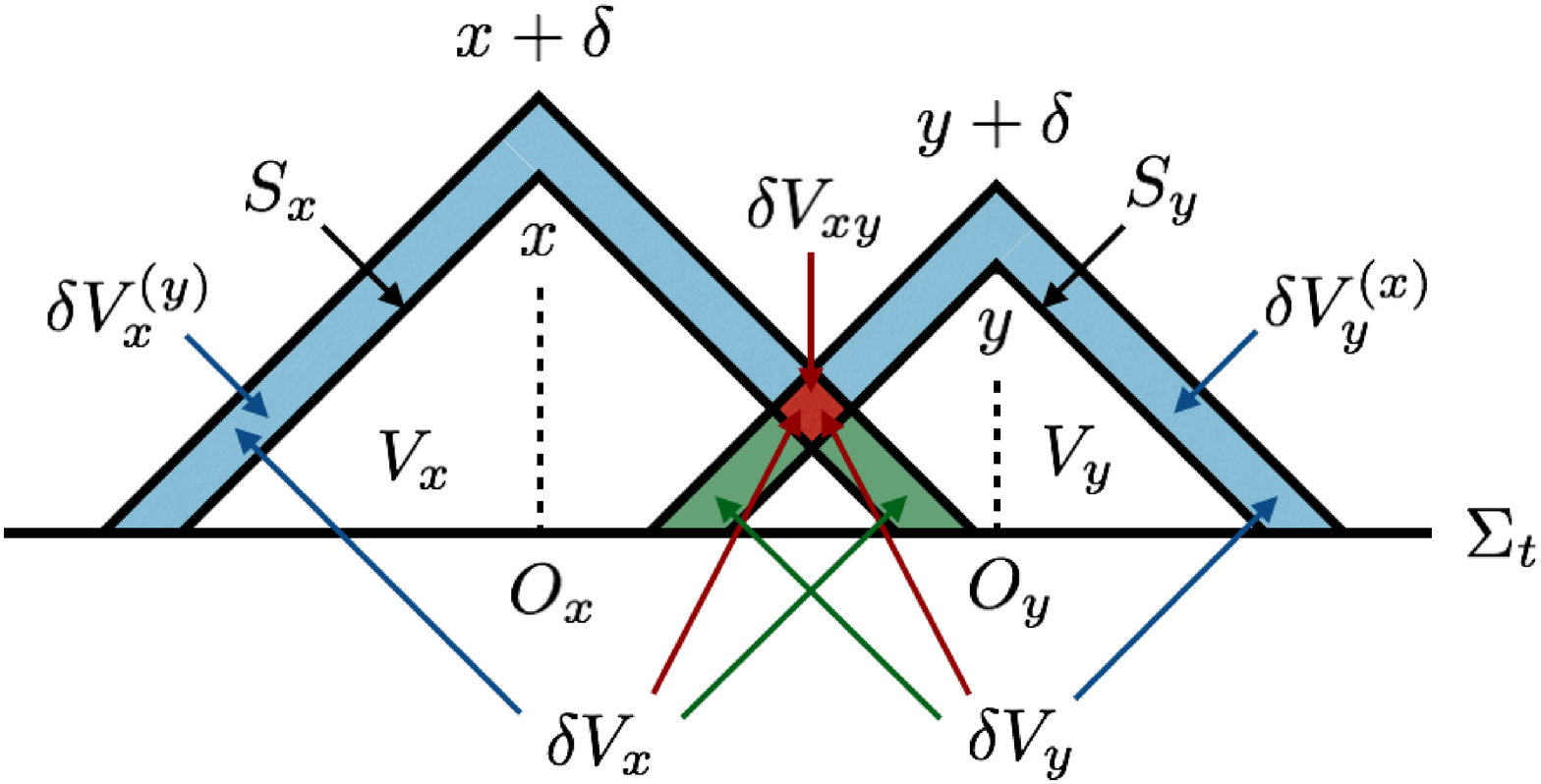}
\caption {\small
Notations for quantities on the past light cones of $x$ and $y$.
}
\label{fig:LightCone}
\end{center}
\end{figure}

\begin{figure}
\begin{center}
\includegraphics[width=0.8\columnwidth]{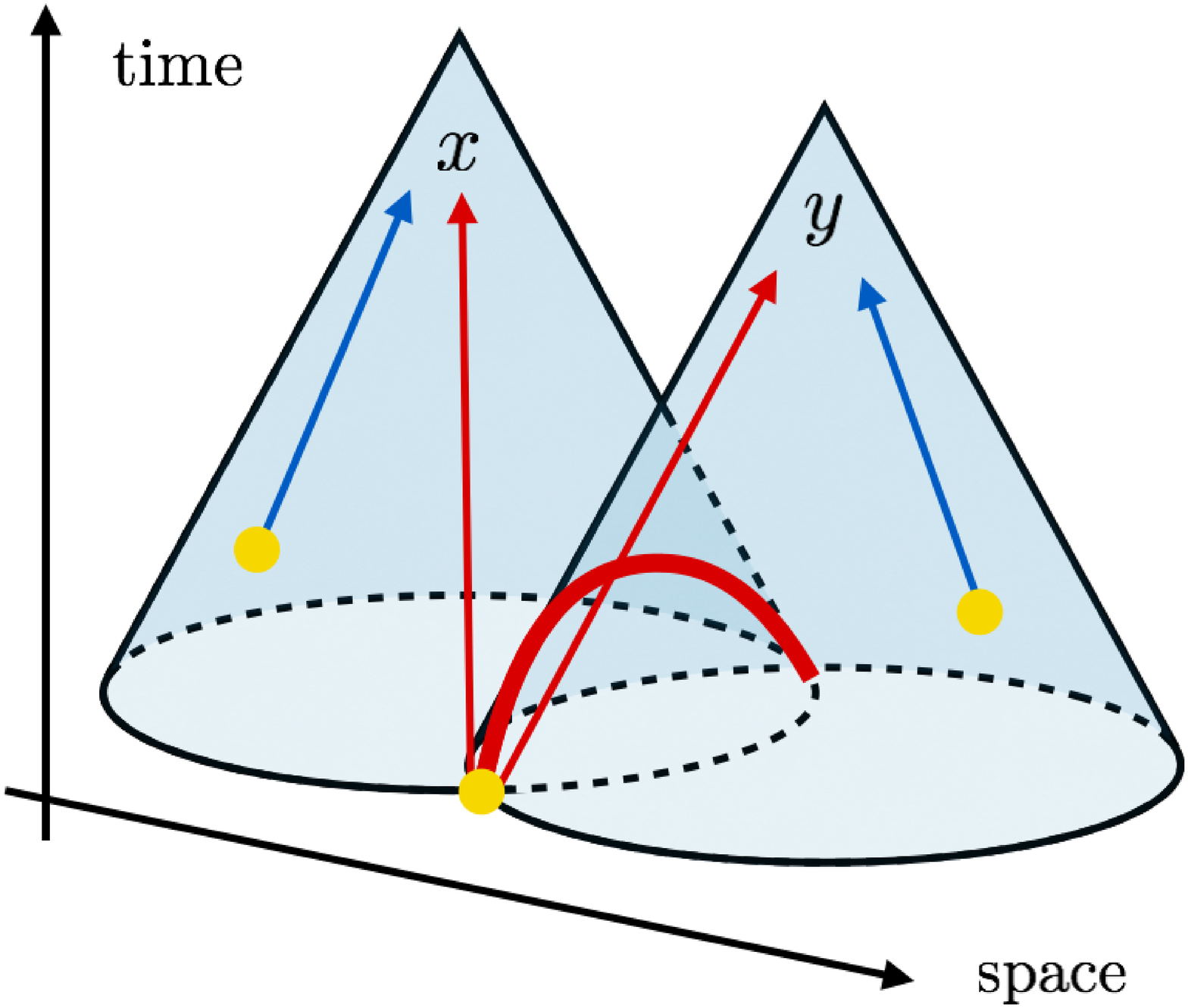}
\caption {\small
How the light cones in Fig.~\ref{fig:LightCone} look like in $2 + 1$ dimensions.
The yellow circles represent the nucleation points for 
single-bubble (on the red central arrows) 
and double-bubble (on the blue separate arrows) 
contributions.
The red line along the intersection of the two light cones shows $\delta V_{xy}$ in Fig.~\ref{fig:LightCone}.
}
\label{fig:LightCone3D}
\end{center}
\end{figure}

We first fix our notations and conventions used in the following argument.
We denote the two spacetime points in the two-point correlator
as (see Fig.~\ref{fig:LightCone} and \ref{fig:LightCone3D})
\begin{align}
x 
&= (t_x,\vec{x}),
\;\;\;
y = (t_y,\vec{y}).
\end{align}
We sometimes use the time variables $(T,t_d)$ defined as
\begin{align}
T
&\equiv \frac{t_x + t_y}{2}, 
\;\;\;
t_d
\equiv t_x - t_y,
\end{align}
instead of $(t_x,t_y)$.
Also, we write their spacial separation as
\begin{align}
\vec{r}
\equiv \vec{x} - \vec{y},
\;\;\;
r
&\equiv |\vec{r}|.
\end{align}
We often consider past light cones of $x$ and $y$,
which are denoted by $S_x$ and $S_y$.
The regions inside $S_x$ and $S_y$ are called $V_x$ and $V_y$, respectively,
and we write their union as $V_{xy} \equiv V_x \cup V_y$.
Since we consider bubbles with wall width $l_B$, 
we also define the spacetime points
\begin{align}
x + \delta 
&\equiv (t_x + l_B,\vec{x}),
\;\;\;
y + \delta \equiv (t_y + l_B,\vec{y}),
\end{align}
whose past light cones are denoted by $S_{x + \delta}$ and $S_{y + \delta}$,
respectively.
We also define the following regions
\begin{align}
\delta V_x
&\equiv V_{x+\delta} - V_x, 
\;\;\;
\delta V_y
\equiv V_{y+\delta} - V_y, 
\end{align}
whose intersection is denoted by
\begin{align}
\delta V_{xy}
&\equiv \delta V_x \cap \delta V_y.
\end{align}
In addition, we define
\begin{align}
\delta V_x^{(y)}
&\equiv \delta V_x - V_{y+\delta},
\;\;\;
\delta V_y^{(x)}
\equiv \delta V_y - V_{x+\delta},
\end{align}
as shown in Fig.~\ref{fig:LightCone}.
Also, in Fig.~\ref{fig:LightCone3D}, 
we show how Fig.~\ref{fig:LightCone} looks in $2 + 1$ dimensions.

\begin{figure}
\begin{center}
\includegraphics[width=\columnwidth]{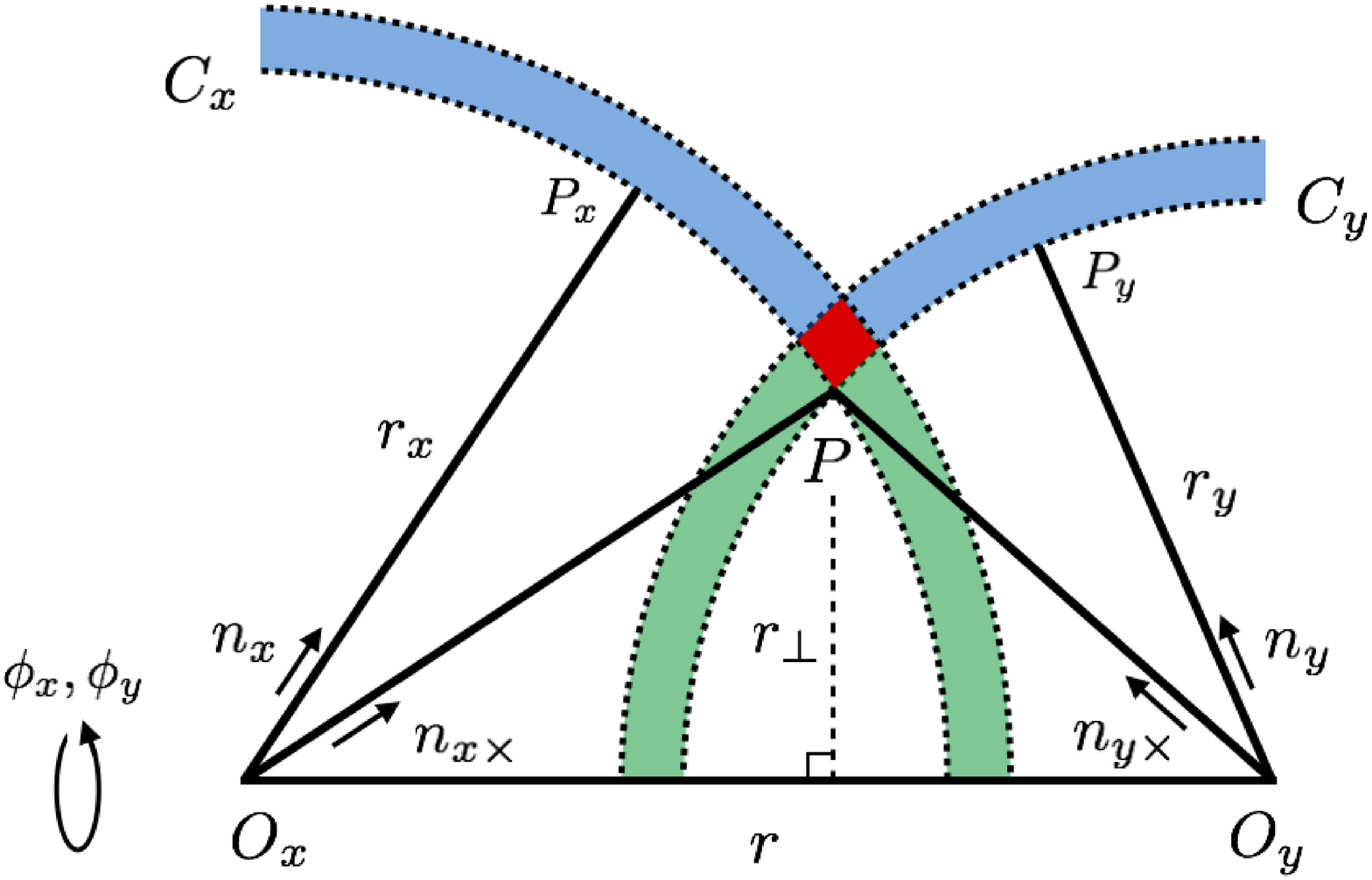}
\caption {\small
Notations for quantities on the constant-time hypersurface $\Sigma_t$.
The red diamond-shaped region denotes the one where 
a bubble nucleate in single-bubble spectrum (see Sec.~\ref{subsec:single}),
while the outer blue regions between the dotted lines denote the ones where
bubbles nucleate in double-bubble spectrum (see Sec.~\ref{subsec:double}).
}
\label{fig:Triangle}
\end{center}
\end{figure}

On a constant-time hypersurface $\Sigma_t$ at time $t$,
the two past light cones $S_x$ and $S_y$ form spheres,
as shown in Fig.~\ref{fig:Triangle}.
We call these two spheres $C_x(t)$ and $C_y(t)$,
whose centers are labelled by $O_x$ and $O_y$, respectively.
The radii of $C_x(t)$ and $C_y(t)$ are given by
\begin{align}
r_x(t)
&\equiv 
t_x - t,
\;\;\;
r_y(t)
\equiv t_y - t.
\end{align}
These spheres $C_x(t)$ and $C_y(t)$ have an intersection 
for time $t < t_{xy}$ with
\begin{align}
t_{xy}
&\equiv \frac{t_x + t_y - r}{2}.
\end{align}
Let us consider arbitrary points $P_x(t)$ on $C_x(t)$
and $P_y(t)$ on $C_y(t)$, 
and we denote unit vectors from $O_x$ and $O_y$ 
to $P_x(t)$ and $P_y(t)$ as $n_x(t)$ and $n_y(t)$,
respectively.
We parameterize these two unit vectors by the azimuthal and polar angles around $\vec{r}$
as
\begin{align}
n_x
&\equiv (s_x c_{\phi x},s_x s_{\phi x},c_x),
\;\;\;
n_y
\equiv (s_y c_{\phi y},s_y s_{\phi y},c_y),
\end{align}
where the label $t$ has been omitted for simplicity.
Also, we use shorthand notations
$c_x (s_x) \equiv \cos \theta_x (\sin \theta_x)$
$c_{\phi x} (s_{\phi x}) \equiv \cos \phi_x (\sin \phi_x)$ etc. in the following.
We sometimes need to label
an arbitrary point on the intersection of $C_x(t)$ and $C_y(t)$.
We denote such point by $P(t)$, and also denote the unit vectors from $O_x$ and $O_y$ to $P(t)$
as $n_{x\times}(t)$ and $n_{y\times}(t)$, respectively.
These unit vectors are parameterized by the azimuthal and polar angles 
$\theta_{x\times}(t)$, $\theta_{y\times}(t)$, $\phi_{x\times}(t)$ and $\phi_{y\times}(t)$ around $\vec{r}$.
Especially, the cosines of the polar angles are given by
\begin{align}
c_{x\times}(t)
=
\cos\theta_{x\times}(t)
&= - \frac{r_x(t)^2+r^2-r_y(t)^2}{2r_x(t)r},
\\
c_{y\times}(t)
=
\cos\theta_{y\times}(t)
&= \frac{r_y(t)^2+r^2-r_x(t)^2}{2r_y(t)r}.
\end{align}

\subsection{False vacuum probability}

\subsubsection{Probability for one point to remain in the false vacuum}

For illustrative purpose, 
we first consider the probability $P(x)$ that a spacetime point $x$ is in the false vacuum.
This occurs if and only if no bubbles are nucleated in $V_x$.
Dividing $V_x$ into infinitesimal four-dimensional regions $dV_x^i$ so that $V_x = \cup_i dV_x^i$,
the probability that no bubbles are nucleated in $dV_x^i$ is given by $(1-\Gamma dV_x^i)$.
Thus $P(x)$ is written as~\cite{Turner:1992tz}
\begin{align}
P(x)
&= \prod_i \left(1 - \Gamma dV_x^i\right)
= e^{-I(x)},
\end{align}
with
\begin{align}
I(x)
&= \int_{V_x}d^4z \; \Gamma(z).
\end{align}

\subsubsection{Probability for two points to remain in the false vacuum}

Next let us consider the probability $P(x,y)$ 
that given two points $x$ and $y$ 
both remain in the false vacuum.
This probability is expressed in the same way as before
\begin{align}
P(x,y)
&= e^{-I(x,y)}, 
\;\;\;
I(x,y)
= \int_{V_{xy}} d^4z \; \Gamma(z).
\end{align}
Below we assume spacelike separation $r > |t_x-t_y|$,
since only such configuration is relevant for the calculation of GW spectrum,
due to the envelope approximation\footnote{
In the envelope approximation,
it is impossible for two spacetime points 
$x = (t_x,\vec{x})$ and $y = (t_y,\vec{y})$ with timelike separation $t_x - t_y > r$
to be on bubble wall(s).
This is because the spacial point $\vec{x}$ is caught up before $t = t_x$
by the bubble wall which passed through $y$.
}.
Then $I(x,y)$ is written as
\begin{align}
&I(x,y)
= I_x^{(y)}+I_y^{(x)}, \\
&I_x^{(y)} 
= \int_{-\infty}^{t_{xy}} dt \; \frac{\pi}{3}r_x(t)^3\Gamma(t)
(2 - c_{x\times}(t))(1 + c_{x\times}(t))^2 \nonumber \\
&\;\;\;\;\;\;\;\;\;\;
+
\int_{t_{xy}}^{t_x} dt \; \frac{4\pi}{3}r_x(t)^3\Gamma(t) \\
&I_y^{(x)}
= I_x^{(y)} |_{x \leftrightarrow y}.
\end{align}
Here we have different integrands for $t \in [-\infty,t_{xy}]$ and otherwise,
because for the former the integrated volume do not form complete spheres.
The time integration can be performed to give
\begin{align}
I(x,y)
&= 8\pi \Gamma(T) {\mathcal I}(t_d,r), 
\label{eq:I}
\\
{\mathcal I}(t_d,r)
&= e^{t_d/2}+e^{-t_d/2}+\frac{t_d^2-(r^2 + 4r)}{4r}e^{-r/2},
\label{eq:calI}
\end{align}
where we have changed 
the variables from $(t_x,t_y)$ to $(T,t_d)$, 
and adopted $\beta = 1$ unit without loss of generality.

\subsection{Single-bubble spectrum}
\label{subsec:single}

We now evaluate the single-bubble contribution to the correlator (\ref{eq:Pi}).
With the envelope approximation,
the following two conditions are required
in order for a single bubble to give nonvanishing energy-momentum tensor 
at both $x$ and $y$:
\begin{itemize}
\item
No bubbles are nucleated in $V_{xy}$.
\item
At least one bubble is nucleated in $\delta V_{xy}$.
\end{itemize}
Note that the last condition reduces to ``Only one bubble is nucleated in $\delta V_{xy}$"
in the thin-wall limit $l_B \to 0$.
Below, we briefly derive the GW spectrum via single-bubble contribution
starting from these two conditions.
The final expression is Eq.~(\ref{eq:Delta_single}), 
and the details of the calculation are summarized in Appendix~\ref{app:det}.

From above considerations,
single-bubble contribution to the energy-momentum tensor 
is factorized in the following way (``$s$" denotes ``single")
\begin{align}
&\langle T_{ij} T_{kl} \rangle^{(s)} (t_x,t_y,\vec{r}) 
\nonumber \\
&\;\;
= 
P(t_x,t_y,r) \int_{-\infty}^{t_{xy}} dt_n \Gamma(t_n) 
{\mathcal T}^{(s)}_{ij,kl}(t,t_x,t_y,\vec{r}),
\label{eq:single_TT}
\end{align}
where ${\mathcal T}^{(s)}_{ij,kl}$ is the value of $T_{ij}(x)T_{kl}(y)$ 
by the wall of the bubble nucleated at time $t_n$ (see Fig.~\ref{fig:LightCone} and \ref{fig:LightCone3D}).
This is calculated as
\begin{align}
{\mathcal T}^{(s)}_{ij,kl}
&= 
\left( \frac{4\pi}{3} r_x(t_n)^3 \cdot \kappa\rho_0 \cdot \frac{1}{4\pi r_x(t_n)^2l_B} \right) 
\nonumber \\
&\;\;\;\;
\times
\left( \frac{4\pi}{3} r_y(t_n)^3 \cdot \kappa\rho_0 \cdot \frac{1}{4\pi r_y(t_n)^2l_B} \right) \nonumber \\
&\;\;\;\;
\times \int_{R_{xy}} d^3z \; (N_{\times}(t_n))_{ijkl},
\end{align}
with $(N_{\times})_{ijkl} \equiv (n_{x\times})_i (n_{x\times})_j (n_{y\times})_k (n_{y\times})_l$.
Here $R_{xy} \equiv \delta V_{xy} \cap \Sigma_{t_n}$ is the ring made by 
rotating the diamond-shape shown in Fig.~\ref{fig:Triangle} around the axis $\vec{r}$.
The integration by the nucleation time $t_n$ in Eq.~(\ref{eq:single_TT}) can be performed explicitly,
and after taking the projection $K$ in Eq.~(\ref{eq:Pi}) into account, we have
\begin{align}
&K_{ij,kl}(\hat{k})K_{ij,mn}(\hat{k})
\langle T_{kl} T_{mn} \rangle^{(s)} (t_x,t_y,\vec{r}) 
\nonumber \\
&= 
\frac{2\pi}{9} \kappa^2\rho_0^2
\; \Gamma(T)
\frac{e^{-r/2}}{r^5} 
P(t_x,t_y,r)
\nonumber \\
&\;\;\;\;
\times
\left[ \frac{1}{2}F_0 + \frac{1}{4}(1 - (\hat{r} \cdot \hat{k})^2)F_1
+ \frac{1}{16}(1 - (\hat{r} \cdot \hat{k})^2)^2F_2 \right],
\label{eq:KKTT_single}
\end{align}
with $F$ functions given by
\begin{align}
F_0
&= 2(r^2 - t_d^2)^2(r^2 + 6r + 12), \\
F_1
&= 2(r^2 - t_d^2)\left[ -r^2(r^3+4r^2 + 12r + 24) \right. \nonumber \\
&\;\;\;\;\;\;\;\;\;\;\;\;\;\;\;\;\;\;\;\;\;
\left. + t_d^2(r^3 + 12r^2 + 60r + 120) \right], 
\\
F_2
&= \frac{1}{2} \left[
r^4(r^4 + 4r^3 + 20r^2 + 72r + 144) \right. 
\nonumber \\
&\;\;\;\;\;\;\;\;
- 2t_d^2r^2(r^4 + 12r^3 + 84r^2 + 360r + 720) \nonumber \\
&\;\;\;\;\;\;\;\;
\left.
+ \; t_d^4 
(r^4 + 20r^3 + 180r^2 + 840r + 1680) \right].
\end{align}
Note that we have changed the time variables from $(t_x,t_y)$
to $(T,t_d)$.
Also note that the correlator has now been successfully expressed analytically.
Performing the integration over the angle between $\vec{r}$ and $\vec{k}$
in Eq.~(\ref{eq:Pi}), we find
\begin{align}
\Pi^{(s)} (t_x,t_y,k)
&= 
\frac{4\pi^2}{9} \kappa^2\rho_0^2 \;
\Gamma(T)
\int_0^\infty dr \; \frac{e^{-r/2}}{r^3}P(t_x,t_y,r) \nonumber \\
&\;\;\;\;\;\;
\times 
\left[ j_0(kr)F_0 + \frac{j_1(kr)}{kr}F_1 + \frac{j_2(kr)}{k^2r^2}F_2 \right].
\label{eq:Pi_single}
\end{align}
Then the integration over $T$ in Eq.~(\ref{eq:Delta}) is performed 
by using the equality $\int_{-\infty}^\infty dY~e^{-Xe^Y+nY}=(n-1)!/X^n$,
and we obtain
\begin{align}
\Delta^{(s)}
&= 
\frac{k^3}{12\pi}
\int_0^\infty dt_d \int_{t_d}^\infty dr \; \frac{e^{-r/2}\cos (kt_d)}{r^3 {\mathcal I}(t_d,r)} \nonumber \\
&\;\;\;\;\;\;\;\;\;\;
\times 
\left[ j_0(kr)F_0 + \frac{j_1(kr)}{kr}F_1 + \frac{j_2(kr)}{k^2r^2}F_2 \right],
\label{eq:Delta_single}
\end{align}
where $j_{0,1,2}$ denote the spherical Bessel functions given in Appendix~\ref{app:det}.

\subsection{Double-bubble spectrum}
\label{subsec:double}

Next we evaluate the double-bubble contribution to the correlator (\ref{eq:Pi}).
With the envelope approximation,
the following two conditions are necessary and sufficient
for two different bubbles to give nonvanishing energy-momentum tensor 
at $x$ and $y$:
\begin{itemize}
\item
No bubbles are nucleated in $V_{xy}$.
\item
At least one bubble is nucleated in $\delta V_x^{(y)}$, 
and at least another is nucleated in $\delta V_y^{(x)}$.
\end{itemize}
Note that the last condition reduces to 
``Only one bubble is nucleated in each of $\delta V_x^{(y)}$ and $\delta V_y^{(x)}$"
in the thin-wall limit $l_B \to 0$.
Below we derive the GW spectrum via double-bubble contribution
starting from these two conditions.
The final result is given by Eq.~(\ref{eq:Delta_double}).

From above considerations, 
the two-bubble contribution to the energy-momentum tensor is 
decomposed as (``$d$" denotes ``double") 
\begin{align}
&\langle T_{ij} T_{kl} \rangle^{(d)} (t_x,t_y,\vec{r}) 
\nonumber \\
&= 
P(t_x,t_y,r) 
\nonumber \\
&\;\;\;\;
\int_{-\infty}^{t_{xy}} dt_{xn}
\Gamma(t_{xn})
\int_{\delta V_x^{(y)} \cap \Sigma_{t_{xn}}} d^3x_n \;
{\mathcal T}^{(d)}_{x,ij}(t_{xn},\vec{x}_n;t_x,\vec{r}) 
\nonumber \\
&\;\;\;\;
\times
\int_{-\infty}^{t_{xy}} dt_{yn}
\Gamma(t_{yn})
\int_{\delta V_y^{(x)} \cap \Sigma_{t_{yn}}} d^3y_n \;
{\mathcal T}^{(d)}_{y,kl}(t_{yn},\vec{y}_n;t_y,\vec{r}),
\label{eq:Double_TT}
\end{align}
where ${\mathcal T}^{(d)}_{x,ij}$ and ${\mathcal T}^{(d)}_{y,kl}$ 
are the value of the energy-momentum tensor 
by the bubble wall nucleated in $\vec{x}_n \in \delta V_x^{(y)} \cap \Sigma_{t_{xn}}$ 
and $\vec{y}_n \in \delta V_y^{(x)} \cap \Sigma_{t_{yn}}$
evaluated at the spacetime points $x$ and $y$, respectively.
They are given by
\begin{align}
&{\mathcal T}^{(d)}_{x,ij}(t_{xn},\vec{x}_n;t_x,\vec{r}) \nonumber \\
&\;\;
= 
\left( \frac{4\pi}{3}r_x(t_{xn})^3 \cdot \kappa\rho_0 \cdot \frac{1}{4\pi r_x(t_{xn})^2l_B} \right) 
(n_x)_i (n_x)_j,
\nonumber \\
&{\mathcal T}^{(d)}_{y,kl}(t_{yn},\vec{y}_n;t_y,\vec{r}) \nonumber \\
&\;\;
= 
\left( \frac{4\pi}{3}r_y(t_{yn})^3 \cdot \kappa\rho_0 \cdot \frac{1}{4\pi r_y(t_{yn})^2l_B} \right)
(n_y)_i (n_y)_j.
\end{align}
Here the arguments $t_{xn}$ and $t_{yn}$ in $n_x$ and $n_y$ are omitted for simplicity.
Note that the time integration is over $[-\infty,t_{xy}]$ in Eq.~(\ref{eq:Double_TT}),
because the integration region $t_{xn} > t_{xy}$ or $t_{yn} > t_{xy}$ 
gives spherically symmetric contribution and thus vanishes 
(see Fig.~\ref{fig:LightCone}--\ref{fig:Triangle}, 
and notice that the nucleation points $P_x$ and $P_y$ run over the whole sphere
for these nucleation times).
Also note that the contribution to $x$ and that to $y$ factorize in Eq.~(\ref{eq:Double_TT})
because the two bubbles nucleate independently of each other 
(see Fig.~\ref{fig:LightCone} and \ref{fig:LightCone3D}).
There are no special directions except for $\vec{r}$, 
and therefore ${\mathcal T}_{z,ij}^{(d)}$ ($z = x,y$) is decomposed as follows 
after integration over the nucleation time $t_{zn}$:
\begin{align}
&\int_{-\infty}^{t_{xy}} dt_{zn} 
\int d^3z_n 
\; 
{\mathcal T}^{(d)}_{z,ij}(t_{zn},\vec{z}_n;t_z,\vec{r}) 
\nonumber \\
&\;\;
= {\mathcal A}^{(d)}_z(t_x,t_y,r) \delta_{ij}
+ {\mathcal B}^{(d)}_z(t_x,t_y,r) \hat{r}_i \hat{r}_j.
\end{align}
Here ${\mathcal A}_z^{(d)}$ and ${\mathcal B}_z^{(d)}$ depend on 
both $t_x$ and $t_y$ because the integration region for $z_n$ 
is affected by the other points.
After the projection by $K$, only ${\mathcal B}$ component survives:
\begin{align}
&K_{ij,kl}(\hat{k})K_{ij,mn}(\hat{k})
\langle T_{kl} T_{mn} \rangle^{(d)} (t_x,t_y,\vec{r}) \nonumber \\
&= \frac{1}{2}
P(t_x,t_y,r) 
{\mathcal B}^{(d)}_x(t_x,t_y,r)
{\mathcal B}^{(d)}_y(t_x,t_y,r)
(1 - (\hat{r} \cdot \hat{k})^2)^2.
\label{eq:KKTT_double}
\end{align}
Taking $\beta = 1$ unit without loss of generality,
we can calculate ${\mathcal B}$ as
\begin{align}
{\mathcal B}^{(d)}_x(t_x,t_y,r)
&= -\frac{\pi}{6} \kappa\rho_0 \; \frac{e^{-r/2}}{r^3} \Gamma(T) G(t_d,r), \\
{\mathcal B}^{(d)}_y(t_x,t_y,r)
&= -\frac{\pi}{6} \kappa\rho_0 \; \frac{e^{-r/2}}{r^3} \Gamma(T) G(-t_d,r),
\end{align} 
with $G$ function given by
\begin{align}
G(t_d,r)
&= (r^2 - t_d^2)\left[ (r^3 + 2r^2) + t_d(r^2 + 6r + 12) \right].
\end{align}
Note that we have now expressed the correlator analytically.
As in the single-bubble case, the angular integration is readily calculated
\begin{align}
&\Pi^{(d)} (t_x,t_y,k) \nonumber \\
&=
\frac{4\pi^3}{9} \kappa^2\rho_0^2
\Gamma(T)^2
\int_0^\infty dr \;
P(t_x,t_y,r)\frac{e^{-r}}{r^4} \nonumber \\
&\;\;\;\;\;\;\;\;\;\;\;\;\;\;\;\;\;\;\;\;\;\;\;\;\;\;\;\;\;\;\;\;
\times \frac{j_2(kr)}{k^2r^2}G(t_d,r)G(-t_d,r).
\label{eq:Pi_double}
\end{align}
Substituting this into Eq.~(\ref{eq:Delta}), and performing $T$ integration
again by using the equality $\int_{-\infty}^\infty dY~e^{-Xe^Y+nY}=(n-1)!/X^n$, 
we have
\begin{align}
\Delta^{(d)}
&= \frac{k^3}{96\pi}
\int_0^\infty dt_d \int_{t_d}^\infty dr \; 
\frac{e^{-r} \cos(k t_d)}{r^4{\mathcal I}(t_d,r)^2} \nonumber \\
&\;\;\;\;\;\;\;\;\;\;\;\;\;\;\;\;\;\;\;\;\;\;\;\;\;\;\;\;
\times \frac{j_2(kr)}{k^2r^2}G(t_d,r)G(-t_d,r).
\label{eq:Delta_double}
\end{align}

\section{Numerical estimation}
\label{sec:num}

Since the remaining integrations 
in Eqs.~(\ref{eq:Delta_single}) and (\ref{eq:Delta_double}) 
cannot be performed explicitly, 
we evaluate them numerically in this section.

\begin{figure}
\begin{center}
\includegraphics[width=\columnwidth]{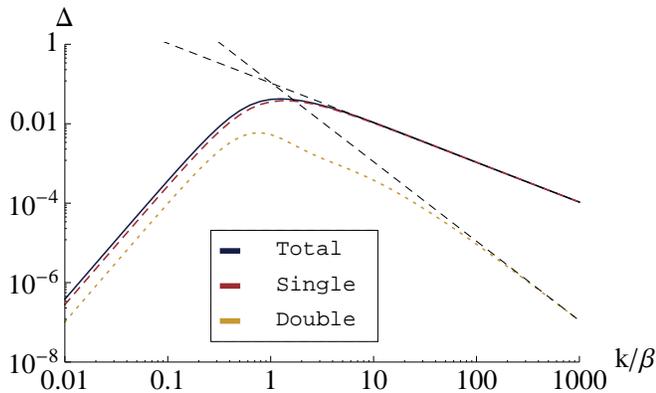}
\caption {\small
Plot of the GW spectrum $\Delta$ (blue).
Single- and double-bubble spectra $\Delta^{(s)}$ (red) and $\Delta^{(d)}$ (yellow)
are also plotted. 
Black lines are auxiliary ones proportional to $k^{-1}$ and $k^{-2}$, respectively.
}
\label{fig:Delta}
\end{center}
\end{figure}

In Fig.~\ref{fig:Delta}, the single- and double-bubble spectra $\Delta^{(s)}$ and $\Delta^{(d)}$
as well as their sum $\Delta = \Delta^{(s)} + \Delta^{(d)}$ are plotted.
As seen from the figure, the low and high frequency behavior is 
\begin{align}
\Delta^{(s)}
&\propto 
\left\{
\begin{array}{ll}
k^3 
\;\;\; & (k/\beta \lesssim 1) \\
k^{-1}
\;\;\; & (1 \lesssim k/\beta)
\end{array}
\right. , \\
\Delta^{(d)}
&\propto 
\left\{
\begin{array}{ll}
k^3 
\;\;\; & (k/\beta \lesssim 1) \\
k^{-1}
\;\;\; & (1 \lesssim k/\beta \lesssim 10) \\
k^{-2}
\;\;\; & (10 \lesssim k/\beta)
\end{array}
\right. .
\end{align}
Since $\Delta^{(s)}$ always dominates $\Delta^{(d)}$,
their sum $\Delta$ behaves as
\begin{align}
\Delta
&\propto 
\left\{
\begin{array}{ll}
k^3 
\;\;\; & (k/\beta \lesssim 1) \\
k^{-1}
\;\;\; & (1 \lesssim k/\beta)
\end{array}
\right. ,
\end{align}
and thus the high-frequency behavior in Ref.~\cite{Huber:2008hg} is confirmed.
Notice that we have restored $\beta$ in the expressions above.

Here we provide a fitting formula for the spectrum
\begin{align}
\Delta 
&= \frac{\Delta_{\rm peak}}
{c_l\left( \frac{f}{f_{\rm peak}} \right)^{-3}
+ (1 - c_l - c_h) \left( \frac{f}{f_{\rm peak}} \right)^{-1}
+ c_h\left( \frac{f}{f_{\rm peak}} \right)},
\label{eq:Delta_fit_v=1}
\end{align}
with
$\Delta_{\rm peak} = 0.043$, $f_{\rm peak}/\beta = 1.24 / 2\pi \simeq 0.20$ 
and $(c_l, c_h) = (0.064,0.48)$.
Here ``$l$" and ``$h$" denote ``low-frequency" and ``high-frequency", respectively,
and note that $f/f_{\rm peak} = k/k_{\rm peak}$.
This formula reproduces the true spectrum within $8\%$ error.
The present spectrum is obtained by substituting 
Eq.~(\ref{eq:Delta_fit_v=1}) into Eq.~(\ref{eq:Omega_present}),
with $f_{\rm peak}$ given by the present value (\ref{eq:f_present}).

\section{Finite velocity}
\label{sec:vel}

Though we have assumed luminal bubble walls in Secs.~\ref{sec:ana}--\ref{sec:num},
we can easily generalize our results to an arbitrary value of the wall velocity.
Just replacing parameters as $r_\bullet \rightarrow r_\bullet/v$ and $k\rightarrow vk$,
we can estimate GWs in almost the same way as in the luminal case.
As a result, we only have to replace the correlation function of the energy-momentum tensor as
\begin{align}
\Pi(t_x,t_y,k)
&\to 
v^3 \Pi(t_x,t_y,vk),
\end{align}
which means that the correlation function for $v \neq c$ is given by
$\Pi = \Pi_v(t_x,t_y,k) = v^3\Pi_c(t_x,t_y,vk)$ with $\Pi_v$ and $\Pi_c$
being the correlation function in $v \neq c$ and $v = c$ case, respectively.
Therefore we obtain
\begin{align}
\Delta
&= \Delta^{(s)} + \Delta^{(d)},
\end{align}
with
\begin{align}
\Delta^{(s)}
&= 
\frac{v^3k^3}{12\pi}
\int_0^\infty dt_d \int_{t_d}^\infty dr \; \frac{e^{-r/2}\cos (kt_d)}{r^3 {\mathcal I}(t_d,r)} \nonumber \\
&\;\;\;\;\;\;\;\;\;\;\;\;
\times 
\left[ j_0(vkr)F_0 + \frac{j_1(vkr)}{vkr}F_1 + \frac{j_2(vkr)}{v^2k^2r^2}F_2 \right],
\label{eq:Delta_single_v} 
\\
\Delta^{(d)}
&= \frac{v^3k^3}{96\pi}
\int_0^\infty dt_d \int_{t_d}^\infty dr \; 
\frac{e^{-r} \cos(k t_d)}{r^4{\mathcal I}(t_d,r)^2} \nonumber \\
&\;\;\;\;\;\;\;\;\;\;\;\;\;\;\;\;\;\;\;\;\;\;\;\;\;\;
\times \frac{j_2(vkr)}{v^2k^2r^2}G(t_d,r)G(-t_d,r).
\label{eq:Delta_double_v}
\end{align}
Note that all the quantities are normalized by $\beta$ in the expressions above.

\begin{figure}
\begin{center}
\includegraphics[width=\columnwidth]{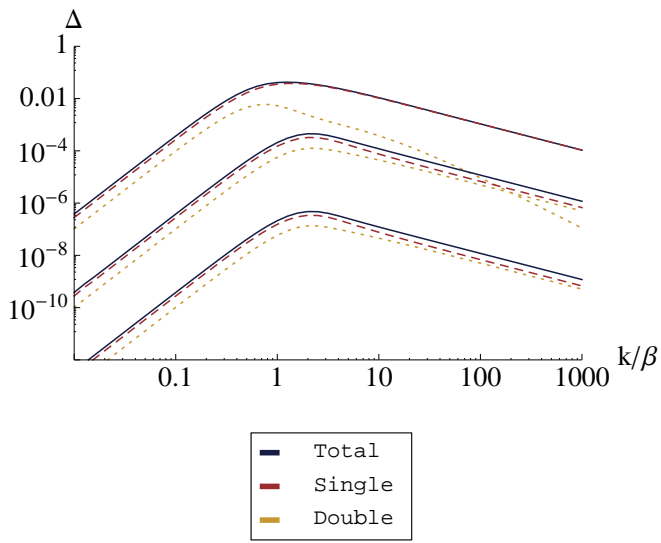}
\caption {\small
Plot of the GW spectrum $\Delta$ (blue)
for $v = 1$, $0.1$ and $0.01$ from top to bottom.
Red and yellow lines correspond to single and double bubble spectrum $\Delta^{(s)}$ and $\Delta^{(d)}$,
respectively.
}
\label{fig:Comparison}
\end{center}
\end{figure}

\begin{figure}
\begin{center}
\includegraphics[width=\columnwidth]{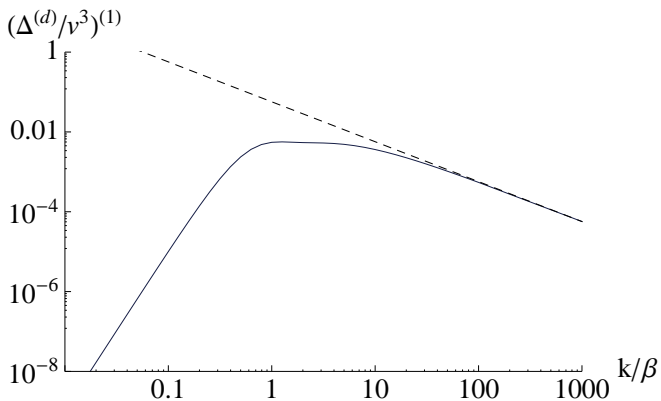}
\caption {\small
Plot of the expansion coefficient $(\Delta^{(d)}/v^3)^{(1)}$ 
in Eq.~(\ref{eq:Delta_expansion}) (blue) 
and an auxiliary line proportional to $k^{-1}$ (black).
This figure shows that $(\Delta^{(d)}/v^3)^{(1)}$ scales as $k^{-1}$ for high frequencies.
}
\label{fig:Expansion}
\end{center}
\end{figure}

As in the luminal case, it is difficult to proceed further in an analytical way,
and hence we perform numerical calculation.
Fig.~\ref{fig:Comparison} is the plot of the GW spectrum $\Delta$ 
for $v = 1$, $0.1$ and $0.01$ from top to bottom.
The single- and double-bubble spectra are also shown in the same figure.
From these plots one sees that $\Delta^{(d)}$ behaves $k^{-2}$ only for $v = c$,
and in other cases $\Delta^{(s)}$ and $\Delta^{(d)}$ both behave as $\propto k^3$ and $\propto k^{-1}$ 
for low and high frequencies, respectively.
This behavior is understood with the following Taylor expansion of $\Delta^{(d)}/v^3$ 
in terms of the wall velocity:
\begin{align}
\frac{\Delta^{(d)}}{v^3}
&= \left( \frac{\Delta^{(d)}}{v^3} \right)^{(0)} 
+ \left(1 - \frac{v}{c} \right)\left( \frac{\Delta^{(d)}}{v^3} \right)^{(1)}
+ \; \cdots
\label{eq:Delta_expansion}
\end{align}
Here note that all the terms vanish except for the first one for $v = c$.
The next-leading term given by
\begin{align}
\left( \frac{\Delta^{(d)}}{v^3} \right)^{(1)}
&= \frac{k^3}{96\pi}
\int_0^\infty dt_d \int_{t_d}^\infty dr \; 
\frac{e^{-r} \cos(k t_d)}{r^4{\mathcal I}(t_d,r)^2} \nonumber \\
&\;\;\;\;\;\;\;\;\;\;\;\;\;\;\;\;\;\;\;\;\;\;\;\;\;\;
\times \frac{j_3(kr)}{k^2r^2}G(t_d,r)G(-t_d,r),
\end{align}
is plotted in Fig.~\ref{fig:Expansion},
and it shows a clear $k^{-1}$ dependence in high frequency region.
This makes $\propto k^{-1}$ behavior in the spectra in Fig.~\ref{fig:Comparison} except for $v = c$.
Our result is consistent with Ref.~\cite{Huber:2008hg} qualitatively,
and also quantitatively within a factor of $2$.

\begin{figure}
\begin{center}
\includegraphics[width=0.9\columnwidth]{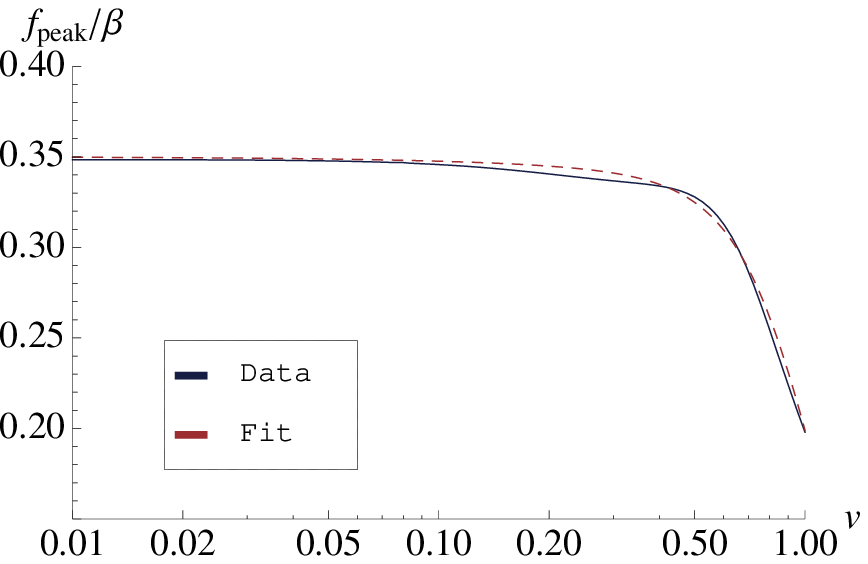}
\caption {\small
Plot of the peak frequency $f_{\rm peak}/\beta$ as a function of the 
bubble wall velocity $v$.
The blue line is numerically calculated from the analytic expression 
(\ref{eq:Delta_single_v}) and (\ref{eq:Delta_double_v}),
while the red line corresponds to the fitting formula (\ref{eq:Fpeak_fit}).
}
\label{fig:Fpeak}
\end{center}
\begin{center}
\includegraphics[width=0.9\columnwidth]{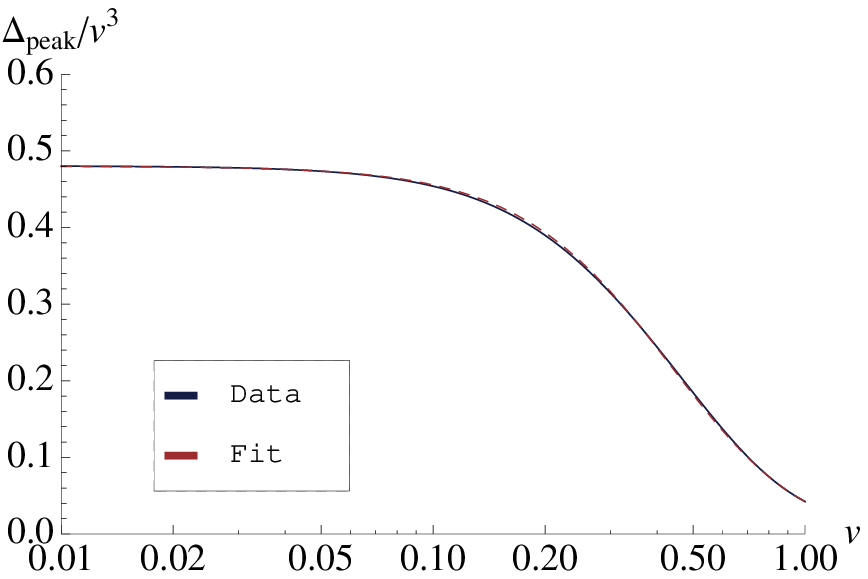}
\caption {\small
Plot of the GW amplitude at the peak $\Delta_{\rm peak}$ scaled by $v^3$.
The blue line is numerically calculated from the analytic expression
(\ref{eq:Delta_single_v}) and (\ref{eq:Delta_double_v}),
while the red line corresponds to the fitting formula (\ref{eq:Deltapeak_fit}).
}
\label{fig:Deltapeak}
\end{center}
\end{figure}

Finally we provide approximate formulae for the frequency and the spectrum at the peak,
as well as the one for the present GW spectrum.
The wall-velocity dependence of the peak frequency and amplitude 
is shown in Figs.~\ref{fig:Fpeak}--\ref{fig:Deltapeak} as blue lines, 
while the red lines are the following fitting formulae:
\begin{align}
&\frac{f_{\rm peak}}{\beta}
= 
\frac{0.35}{1 + 0.069~ v + 0.69~ v^4}, 
\label{eq:Fpeak_fit}
\\
&\Delta_{\rm peak}
= 
\frac{0.48v^3}{1 + 5.3~ v^2 + 5.0~ v^4},
\label{eq:Deltapeak_fit}
\end{align}
which reproduce the true spectrum within $5\%$ and $3\%$ errors, respectively.
The present peak frequency and amplitude are obtained 
by using Eqs.~(\ref{eq:f_present})--(\ref{eq:Omega_present}),
which are shown here again
\begin{align}
&f
= 1.65 \times 10^{-5} {\rm Hz}
\left( \frac{f_{\rm peak}}{\beta} \right) \left( \frac{\beta}{H_*} \right)
\left( \frac{T_*}{10^2{\rm GeV}} \right)
\left( \frac{g_*}{100} \right)^{\frac{1}{6}}, 
\label{eq:f_present_2} 
\\
&\Omega_{\rm GW}h^2 \nonumber \\
&=1.67\times 10^{-5}\kappa^2 \Delta_{\rm peak} \left( \frac{\beta}{H_*} \right)^{-2}
\left( \frac{\alpha}{1 + \alpha} \right)^2
\left( \frac{g_*}{100} \right)^{-\frac{1}{3}},
\label{eq:Omega_present_2}
\end{align}
where $H_*$ and $T_*$ are the Hubble parameter at the transition 
and the temperature of the universe just after the transition, respectively,
$g_*$ is the number of relativistic degrees of freedom at temperature $T_*$,
$\kappa$ is the efficiency factor defined in Eq.~(\ref{eq:rho}),
and $\alpha$ and $\beta$ are 
the fraction of the released energy density and the parameter in the nucleation rate
defined in Eqs.~(\ref{eq:alpha}) and (\ref{eq:Gamma_beta}), respectively.

\section{Discussion and Conclusions}
\label{sec:con}

In this paper, we have derived analytical expressions 
for the gravitational wave (GW) spectrum from bubble collision 
during cosmological first-order phase transition,
with thin-wall and envelope approximations in a flat background.
(see Eqs.~(\ref{eq:Delta_single}) and (\ref{eq:Delta_double})).
The point is that 
we have only to know the two-point correlator of the energy-momentum tensor 
$\langle T(x)T(y)\rangle$,
which in fact can be expressed in an analytic way.
As a result, it is found that the most of the contributions to the spectrum come from
single-bubble contribution to the correlator,
and in addition the fall-off of the spectrum at high frequencies is found to be proportional to $f^{-1}$.
We have also provided some fitting formulae for the spectrum 
(Eq.~(\ref{eq:Delta_fit_v=1}) and Eqs.~(\ref{eq:Fpeak_fit})--(\ref{eq:Omega_present_2})).

The key assumption which makes the analytic formulae 
quite simple is the thin-wall approximation,
because this assumption enables us to classify various contributions to 
$\langle T(x)T(y)\rangle$ just as ``single-bubble" and ``double-bubble" in Sec.~\ref{sec:ana}.
Therefore, it will be possible to extend our method
to more general setups as long as we adopt the thin-wall approximation.
For example, it may be possible 
to consider more general bubble nucleation rate or to include expansion of the universe.
In addition, it is possible to calculate the GW spectrum analytically 
without the envelope approximation~\cite{JT}.
We leave such studies as future work.

\section*{Acknowledgments}

The work of RJ and MT is supported 
by JSPS Research Fellowships for Young Scientists.

\appendix

\section{Detailed calculation of the single-bubble spectrum}
\label{app:det}

In this appendix we show a detailed calculation of the single-bubble spectrum.
The goal is to derive Eq.~(\ref{eq:Delta_single}).
The quantity we would like to calculate is
\begin{align}
\Delta
&= \frac{3}{8\pi G}\frac{\beta^2\rho_{\rm tot}}{\kappa^2\rho_0^2}\Omega_{\rm GW}\nonumber \\
&= \frac{3}{4\pi^2}\frac{\beta^2k^3}{\kappa^2\rho_0^2}
\int dt_x \int dt_y \cos(k(t_x - t_y))\Pi(t_x,t_y,k).
\label{eq:app_Delta}
\end{align}
Below we take $\beta =1$ unit without loss of generality.
The function $\Pi$ is given by
\begin{align}
&\Pi(t_x,t_y,k) \nonumber \\
&= K_{ij,kl}(\hat{k})K_{ij,mn}(\hat{k})
\int d^3r \; e^{i \vec{k} \cdot \vec{r}} \langle T_{kl} T_{mn} \rangle (t_x,t_y,\vec{r}),
\label{eq:app_Pi}
\end{align}
where $K_{ij,kl}$ denotes the projection given in Eq.~(\ref{eq:kk}).
The single-bubble contribution to the two-point correlator of the energy-momentum tensor 
is decomposed as
\begin{align}
&\langle T_{ij} T_{kl} \rangle^{(s)} (t_x,t_y,\vec{r}) 
\nonumber \\
&\;\;
= 
P(t_x,t_y,r) \int_{-\infty}^{t_{xy}} dt_n \Gamma(t_n) 
{\mathcal T}^{(s)}_{ij,kl}(t,t_x,t_y,\vec{r}),
\end{align}
where ${\mathcal T}^{(s)}_{ij,kl}$ is the value of $T_{ij}(x)T_{kl}(y)$ 
by the wall of the bubble nucleated at time $t_n$ (see Fig.~\ref{fig:LightCone} and \ref{fig:LightCone3D}).
This is calculated as
\begin{align}
{\mathcal T}^{(s)}_{ij,kl}
&= 
\left( \frac{4\pi}{3} r_x(t_n)^3 \cdot \kappa\rho_0 \cdot \frac{1}{4\pi r_x(t_n)^2l_B} \right) 
\nonumber \\
&\;\;\;\;
\times
\left( \frac{4\pi}{3} r_y(t_n)^3 \cdot \kappa\rho_0 \cdot \frac{1}{4\pi r_y(t_n)^2l_B} \right) \nonumber \\
&\;\;\;\;
\times \int_{R_{xy}} d^3z \; (N_{\times})_{ijkl},
\end{align}
with $(N_{\times})_{ijkl} \equiv (n_{x\times})_i (n_{x\times})_j (n_{y\times})_k (n_{y\times})_l$.
Here $R_{xy} \equiv \delta V_{xy} \cap \Sigma_{t_n}$ is the ring made by 
rotating the diamond-shape shown in Fig.~\ref{fig:Triangle} around the axis $\vec{r}$.
In the following we omit the argument $t_n$ in $r_x(t_n)$ and $r_y(t_n)$.
Taking the projection operator $K$ into account, 
and noting that the area of the diamond in Fig.~\ref{fig:Triangle} is $l_B^2/\sin(\theta_x - \theta_y)$, 
we have
\begin{align}
&K_{ij,kl}(\hat{k})K_{ij,mn}(\hat{k}) 
\langle T_{kl}T_{mn} \rangle^{(s)} (t_x,t_y,\vec{r}) \nonumber \\
&= 
\left( \frac{\kappa\rho_0}{3} \right)^2
P(t_x,t_y,\vec{r}) 
\nonumber \\
&\;\;\;\;
\times
\int_{-\infty}^{t_{xy}} dt_n \int d\phi \;
\Gamma(t_n)
\frac{r_x^2r_y^2}{r}
K_{kl,mn} (N_{\times})_{klmn},
\end{align}
Here we have used $r_xs_{x\times}= r_ys_{y\times}$ ($= r_\perp$ in Fig.~\ref{fig:Triangle}), 
$- r_xc_{x\times} + r_yc_{y\times} = r$ 
and $K_{ij,kl}K_{ij,mn} = K_{kl,mn}$.
Also, $\phi$ ($= \phi_{x\times} = \phi_{y\times}$ in Fig.~\ref{fig:Triangle}) 
is the azimuthal angle around $\vec{r}$.
Since there is no special direction except for $\vec{r}$,
the correlator $\langle T_{ij}T_{kl} \rangle^{(s)}$ has only the following terms
\begin{align}
&\langle T_{ij}T_{kl} \rangle^{(s)} \nonumber \\
&= a_1 \delta_{ij} \delta_{kl}
+ a_2 \frac{1}{2}(\delta_{ik}\delta_{jl} + \delta_{il}\delta_{jk}) 
+ b_1\delta_{ij}\hat{r}_k\hat{r}_l
+ b_2\delta_{kl}\hat{r}_i\hat{r}_j \nonumber \\
&\;\;\;\;+ b_3 \frac{1}{4}(\delta_{ik}\hat{r}_j\hat{r}_l + \delta_{il}\hat{r}_j\hat{r}_k
+ \delta_{jk}\hat{r}_i\hat{r}_l + \delta_{jl}\hat{r}_i\hat{r}_k) \nonumber \\
&\;\;\;\;
+c_1 \hat{r}_i\hat{r}_j\hat{r}_k\hat{r}_l,
\end{align}
with $a$, $b$ and $c$ denoting some coefficients.
After projection, only a few terms survive:
\begin{align}
&K_{ij,kl}(\hat{k})K_{ij,mn}(\hat{k})
\langle T_{kl}T_{mn} \rangle^{(s)} \nonumber \\
&\;\;\;\;\;\;\;\;\;\;\;\;
= 2a_2 
+ (1-c_{rk}^2)b_3 
+ \frac{1}{2}(1-c_{rk}^2)^2c_1,
\end{align}
with $c_{rk} \equiv \hat{r} \cdot \hat{k}$.
Coefficients $a,b,c$ can be extracted by identifying $\vec{r}$ as $z$ direction
\begin{align}
&\langle T_{xx}T_{xx} \rangle^{(s)}
= a_1 + a_2, 
\;\;\;
\langle T_{xy}T_{xy} \rangle^{(s)}
= \frac{1}{2}a_2, \nonumber \\
&\langle T_{xx}T_{zz} \rangle^{(s)}
= a_1 + b_1,
\;\;\;
\langle T_{zz}T_{xx} \rangle^{(s)}
= a_1 + b_2, \nonumber \\
&\langle T_{xz}T_{xz} \rangle^{(s)}
= \frac{1}{2}a_2 + \frac{1}{4}b_3, 
\nonumber \\
&\langle T_{zz}T_{zz} \rangle^{(s)}
= a_1 + a_2 + b_1 + b_2 + b_3 +c_1,
\end{align}
which give
\begin{align}
a_2 
&= 2\langle T_{xy}T_{xy} \rangle^{(s)}, \nonumber \\
b_3
&= 4(\langle T_{xz}T_{xz} \rangle^{(s)} - \langle T_{xy}T_{xy} \rangle^{(s)}), \nonumber \\
c_1
&= \langle T_{xx}T_{xx} \rangle^{(s)}
- (\langle T_{xx}T_{zz} \rangle^{(s)} + \langle T_{zz}T_{xx} \rangle^{(s)}) \nonumber \\
&\;\;\;\;
- 4\langle T_{xz}T_{xz} \rangle^{(s)} 
+ \langle T_{zz}T_{zz} \rangle^{(s)}.
\end{align}
Therefore, we can write down the projected correlator as 
\begin{align}
&K_{ij,kl}(\hat{k})K_{ij,mn}(\hat{k}) 
\langle T_{kl}T_{mn} \rangle^{(s)} (t_x,t_y,\vec{r}) \nonumber \\
&= 
\left( \frac{\kappa\rho_0}{3} \right)^2
\;
P(t_x,t_y,r)
\int_{-\infty}^{t_{xy}} dt_n \int d\phi \; 
\Gamma(t_n)
\frac{r_x^2r_y^2}{r} 
\nonumber \\
&\;\;\;\;
\biggl[ 
4N_{xy,xy}
+ 4(1 - c_{rk}^2)(N_{xz,xz} - N_{xy,xy}) + \frac{1}{2}(1 - c_{rk}^2)^2
\nonumber \\
&\;\;\;\;\;\;
\times
(N_{xx,xx}
- (N_{xx,zz} + N_{zz,xx})
- 4N_{xz,xz} + N_{zz,zz}) 
\biggr]
\nonumber \\
&=
\frac{2\pi}{9}\kappa^2\rho_0^2
\;
P(t_x,t_y,r)
\int_{-\infty}^{t_{xy}} dt_n \; 
\Gamma(t_n)
\frac{r_x^2r_y^2}{r} 
\nonumber \\
&\;\;\;\;
\biggl[ 
\frac{1}{2}F'_0
+ \frac{1}{4}(1-c_{rk}^2)F'_1
+ \frac{1}{16}(1-c_{rk}^2)^2F'_2
\biggr].
\end{align}
with
\begin{align}
F'_0
&= 
s_{x\times}^2s_{y\times}^2, \\
F'_1
&= 
8s_{x\times}c_{x\times}s_{y\times}c_{y\times} - 2s_{x\times}^2s_{y\times}^2, \\
F'_2
&= 
3s_{x\times}^2s_{y\times}^2 - 4(s_{x\times}^2c_{y\times}^2 + c_{x\times}^2s_{y\times}^2)
\nonumber \\
&\;\;\;\;
- 16s_{x\times}c_{x\times}s_{y\times}c_{y\times} + 8c_{x\times}^2c_{y\times}^2.
\end{align}
Using $r_xs_{y\times} = r_ys_{y\times}$, 
we may arrange the expressions in the square parenthesis 
so that $s_x$ and $s_y$ appear only in $s_x^2$ and $s_y^2$:
\begin{align}
&K_{ij,kl}(\hat{k})K_{ij,mn}(\hat{k}) 
\langle T_{kl}T_{mn} \rangle^{(s)} (t_x,t_y,\vec{r}) \nonumber \\
&=
\frac{2\pi}{9} \kappa^2\rho_0^2 
\; 
P(t_x,t_y,r)
\int_{-\infty}^{t_{xy}} dt_n \; 
\frac{\Gamma(t_n)}{r}
\nonumber \\
&\;\;\;\;
\times
\biggl[ 
\frac{1}{2}F''_0
+ \frac{1}{4}(1 - c_{rk}^2)F''_1
+ \frac{1}{16}(1 - c_{rk}^2)^2F''_2 
\biggr],
\end{align}
with
\begin{align}
F''_0
&= r_x^2r_y^2s_{x\times}^2s_{y\times}^2, \\
F''_1
&= 
r_xr_y \left[ 
4c_{x\times}c_{y\times}(r_x^2s_{x\times}^2 + r_y^2s_{y\times}^2)
- 2r_xr_ys_{x\times}^2s_{y\times}^2 
\right], \\
F''_2
&= r_xr_y \bigl[ r_xr_y(19c_{x\times}^2c_{y\times}^2 - 7(c_{x\times}^2 + c_{y\times}^2) + 3) \nonumber \\ 
&\;\;\;\;\;\;\;\;\;\;\;\;\;\;\;\;\;\;\;\;\;\;\;\;\;\;\;\;\;\;\;\;\;\;\;\;
- 8c_{x\times}c_{y\times}(r_x^2s_{x\times}^2 + r_y^2s_{y\times}^2) \bigr].
\end{align}
This allows us to express the integrand without square roots coming from 
$s_{x\times} = \sqrt{1 - c_{x\times}^2}$ etc.

Now the integration by the nucleation time $t_n$ can be performed explicitly
by changing the integration variable from $t$ to $t_T \equiv t_n - T$.
Here notice that $t_T$ integration is from $-\infty$ to $-r/2$ since $t_{xy} = T - r/2$. 
Also note 
that $F''_0$, $F''_1$ and $F''_2$ are polynomials in $t_T$
and that $\Gamma(t_n)$ can be factorized as $\Gamma(t_n) = \Gamma(T)e^{t_T}$.
As a result, we obtain
\begin{align}
&
K_{ij,kl}(\hat{k})K_{ij,mn}(\hat{k}) 
\langle T_{kl}T_{mn} \rangle^{(s)} (t_x,t_y,\vec{r}) \nonumber \\
&= 
\frac{2\pi}{9} \kappa^2\rho_0^2
\; 
P(t_x,t_y,r)
\Gamma(T)\frac{e^{-r/2}}{r^5} \nonumber \\
&\;\;\;\;
\times
\left[ \frac{1}{2}F_0 + \frac{1}{4}(1 - c_{rk}^2)F_1
+ \frac{1}{16}(1 - c_{rk}^2)^2F_2 \right],
\end{align}
with $F$ functions given by 
\begin{align}
F_0
&= 2(r^2 - t_d^2)^2(r^2 + 6r + 12), \\
F_1
&= 2(r^2 - t_d^2)\left[ -r^2(r^3+4r^2 + 12r + 24) \right. \nonumber \\
&\;\;\;\;\;\;\;\;\;\;\;\;\;\;\;\;\;\;\;\;\;
\left. + t_d^2(r^3 + 12r^2 + 60r + 120) \right], \\
F_2
&= \frac{1}{2} \left[
r^4(r^4 + 4r^3 + 20r^2 + 72r + 144) \right. 
\nonumber \\
&\;\;\;\;\;\;\;\;
- 2t_d^2r^2(r^4 + 12r^3 + 84r^2 + 360r + 720) \nonumber \\
&\;\;\;\;\;\;\;\;
\left.
+ t_d^4(r^4 + 20r^3 + 180r^2 + 840r + 1680) \right].
\end{align}

In Eq.~(\ref{eq:app_Pi}) the integration by the angle between 
$\hat{r}$ and $\hat{k}$ is easily calculated,
since the angular dependence appears only through $c_{rk}$.
Noting that
\begin{align}
&\int_{-1}^1 dc \; e^{icx}
= 2j_0(x),
\;\;\;\;
\int_{-1}^1 dc \; e^{icx}(1-c^2)
= \frac{4j_1(x)}{x},
\nonumber \\ 
&\int_{-1}^1 dc \; e^{icx}(1-c^2)^2
= \frac{16j_2(x)}{x^2},
\end{align}
with $j_i$ being the spherical Bessel functions 
\begin{align}
&j_0(x)
= \frac{\sin x}{x}, \nonumber \\
&j_1(x)
= \frac{\sin x - x\cos x}{x^2}, \nonumber \\
&j_2(x)
= \frac{(3-x^2)\sin x - 3x\cos x}{x^3}, \nonumber \\
&j_3(x)
= \frac{(15-6x^2)\sin x - (15x-x^3)\cos x}{x^4},
\end{align}
we have
\begin{align}
\Pi^{(s)} (t_x,t_y,k)
&= 
\frac{4\pi^2}{9} \kappa^2\rho_0^2
\;
\Gamma(T)
\int_0^\infty dr \; \frac{e^{-r/2}}{r^3} P(t_x,t_y,r) \nonumber \\
&\;\;\;\;\;\;
\times 
\left[ j_0(kr)F_0 + \frac{j_1(kr)}{kr}F_1 + \frac{j_2(kr)}{k^2r^2}F_2 \right].
\end{align}

Finally, let us consider the integration with respect to $T \equiv (t_1+t_2)/2$
in Eq.~(\ref{eq:app_Delta}).
Since $P(x,y)$ is given by (see Eq.~(\ref{eq:I}))
\begin{align}
P(x,y)
&=e^{-8\pi \Gamma(T) {\mathcal I}(t_d,r)},
\end{align}
$T$ dependence of $\Pi^{(s)}(t_x,t_y,k)$ appears through the combination
$\Gamma(T)e^{-8\pi\Gamma(T){\mathcal I}(t_d,r)}$.
By using the equality $\int_{-\infty}^\infty dY~e^{-Xe^Y+nY}=(n-1)!/X^n$,
the integration with respect to $T$ is performed analytically.
After all, we have
\begin{align}
\Delta^{(s)}
&= 
\frac{k^3}{12\pi}
\int_0^\infty dt_d \int_{t_d}^\infty dr \; \frac{e^{-r/2}\cos (kt_d)}{r^3 {\mathcal I}(t_d,r)} \nonumber \\
&\;\;\;\;\;\;\;\;\;\;
\times 
\left[ j_0(kr)F_0 + \frac{j_1(kr)}{kr}F_1 + \frac{j_2(kr)}{k^2r^2}F_2 \right].
\end{align}

\section{Comment on ``spherical symmetry"}
\label{app:Saru}

It is well known that spherically symmetric objects do not radiate gravitational waves.
In this appendix we explain why this ``spherical symmetry"
argument do not undermine our calculation, especially the single-bubble contribution.

As is obvious from the derivation, our formalism, especially the single-bubble contribution, 
does not mean that the two evaluation points $\vec{x}$ and $\vec{y}$ 
are summed over the surface of a sphere. 
Instead, we first fix these evaluation points, 
and then sum over all the bubble configurations 
where the two bubble wall fragments originate from a single nucleation point. 
The very process of fixing $x$ and $y$ automatically takes into account 
the breaking of the spherical symmetry, because the parts of the bubble wall propagating towards the evaluation points are required to be uncollided until they reach these points, while no condition is imposed on other parts of the wall.

\begin{figure}
\begin{center}
\includegraphics[width=0.7\columnwidth]{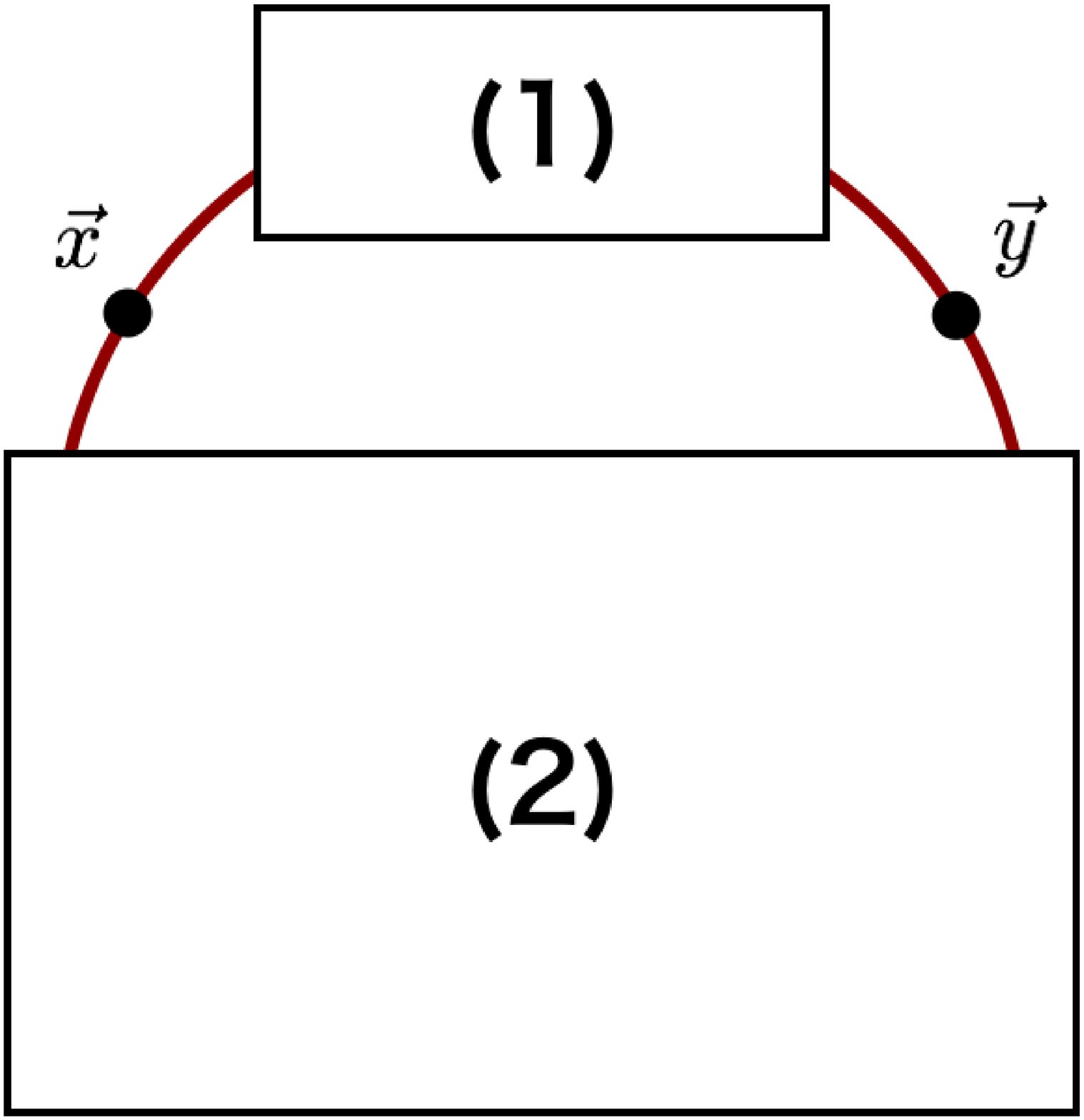}
\\
\vspace{5mm}
\includegraphics[width=\columnwidth]{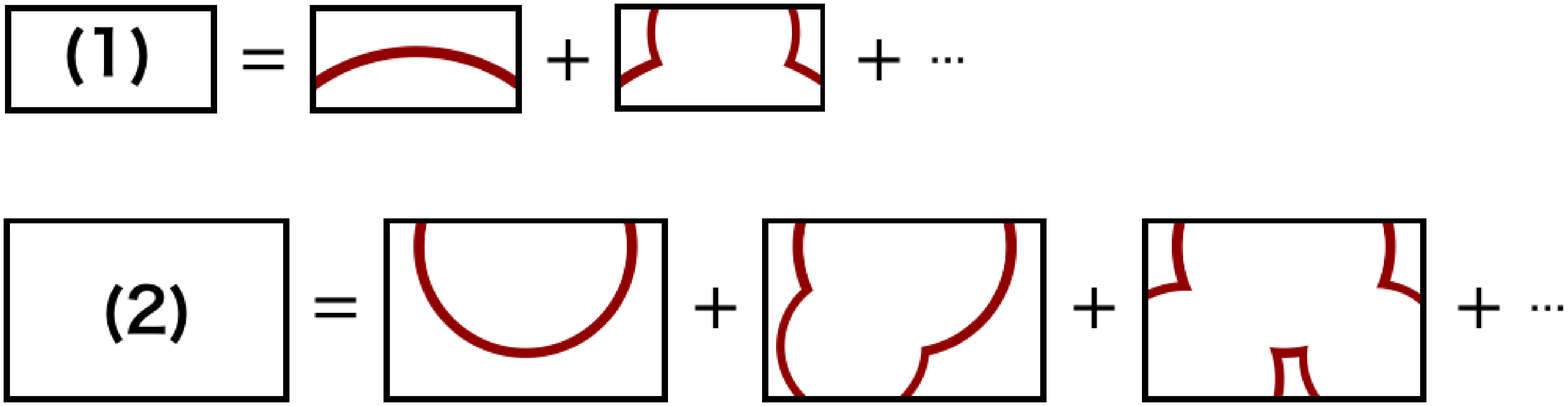}
\caption {\small
Illustration of why the “spherical symmetry” of a single bubble do not undermine our argument. 
We set the evaluation time to be $t_x = t_y$ in this figure.
We require the bubble walls fragments propagating to $\vec{x}$ and $\vec{y}$ 
to be uncollided until the evaluation time,
while other parts of this bubble can be already collided with others.
This automatically takes into account the breaking of the spherical symmetry of a single bubble.
}
\label{fig:Saru}
\end{center}
\end{figure}

We illustrate this point in Fig.~\ref{fig:Saru}. 
Suppose we fix $x$ and $y$ so that $t_x = t_y$. 
Then, at this evaluation time, 
bubble walls in (1) and (2) regions remain uncollided and form part of a complete sphere 
in some cases (first terms in the R.H.S. of the equations), 
while in other cases they are already collided with other walls
(the other terms in the same equations). 
On the other hand, the bubble wall fragments propagating towards $\vec{x}$ and $\vec{y}$ 
must remain uncollided until the evaluation time. 
Therefore, our formalism has nothing to do with the ``spherical symmetry" argument,
and automatically takes account of the breaking of the spherical symmetry by bubble collisions.

\end{document}